\pdfoutput=1

\documentclass[letterpaper]{nature}

\usepackage{booktabs}
\usepackage{lineno}
\usepackage{lscape}
\usepackage{makecell}
\usepackage{ccaption}
\usepackage{amsfonts}
\usepackage{wrapfig}
\usepackage{listings}
\usepackage[utf8]{inputenc}
\usepackage[left=1in,right=1in,bottom=1.1in,top=1in]{geometry}
\usepackage{amsmath}
\usepackage{mathtools}
\usepackage{amssymb}
\usepackage[algoruled]{algorithm2e}
\usepackage{color}
\usepackage{enumitem}
\usepackage{lscape}
\usepackage{dsfont}
\usepackage{booktabs}
\usepackage{amsmath}
\usepackage{amsfonts}
\usepackage[table,xcdraw,dvipsnames]{xcolor}
\usepackage[normalem]{ulem}
\usepackage{float}
\usepackage[colorlinks,citecolor=blue,urlcolor=magenta,linkcolor=blue]{hyperref}
\usepackage{multirow}
\usepackage{xspace}
\usepackage{graphicx}
\usepackage{pbox}
\usepackage{chngpage}
\usepackage{url}
\usepackage{booktabs} 
\usepackage{times} 
\usepackage[innercaption]{sidecap}
\usepackage[font=footnotesize,labelfont=bf]{caption}
\usepackage[CaptionAfterwards]{fltpage}
\usepackage{lineno}
\usepackage{subcaption}
\usepackage{placeins}
\usepackage{verbatim}
\usepackage{cite}
 \usepackage{soul}

\makeatletter
\let\LN@align\align
\let\LN@endalign\endalign
\renewcommand{\align}{\linenomath\LN@align}
\renewcommand{\endalign}{\LN@endalign\endlinenomath}
\makeatother

\newcommand\trackchange[1]{{#1}}

\title{\large 
    \begin{center}
        Quantifying disparities in intimate partner violence: \\ a machine learning method to correct for underreporting
    \end{center}
}

\author{\begin{center}
\vspace{-5mm}
Divya Shanmugam$^{1,5}$, Kaihua Hou$^{2}$, and Emma Pierson$^{3,4}$\\[5mm]
\footnotesize{$^{1}\;$Department of Electrical Engineering and Computer Science, MIT, Cambridge, MA 02139, USA} \\[1mm]
\footnotesize{$^{2}\;$Malone Center of Engineering in Healthcare, Johns Hopkins University School of Medicine, Baltimore, Maryland} \\[1mm]
\footnotesize{$^{3}\;$Department of Computer Science, Cornell Tech, New York, NY 10044, USA} \\[1mm]
\footnotesize{$^{4}\;$Department of Population Health Sciences, Weill Cornell Medical College, New York, NY 10021, USA} \\[1mm] 
\footnotesize{$^{5}\;$Indicates corresponding author. Please address all correspondence to divyas@mit.edu.} \\[15mm] 
\end{center}}

\usepackage[parfill]{parskip}

\begin{document}
\maketitle
\begin{abstract}
\section*{Abstract}
\linespread{1.0}\selectfont
\vspace{1em}
The first step towards reducing the pervasive disparities in women's health is to quantify them. Accurate estimates of the \emph{relative prevalence} across groups --- capturing, for example, that a condition affects Black women more frequently than white women --- facilitate effective and equitable health policy which prioritizes groups who are disproportionately affected by a condition. However, it is difficult to estimate relative prevalence when a health condition is underreported, as many women's health conditions are. In this work, we present \texttt{PURPLE}, a method for accurately estimating the relative prevalence of underreported health conditions which builds upon the positive unlabeled machine learning framework. We show that under a commonly made assumption---that the probability of having a health condition given a set of symptoms remains constant across groups---we can recover the relative prevalence, even without restrictive assumptions commonly made in positive unlabeled learning and even if it is impossible to recover the absolute prevalence. We conduct experiments on synthetic and real health data which demonstrate \texttt{PURPLE}'s ability to recover the relative prevalence more accurately than do previous methods. We then use \texttt{PURPLE} to quantify the relative prevalence of intimate partner violence (IPV) in two large emergency department datasets. 
We find higher prevalences of IPV among patients who are on Medicaid, not legally married, in lower-income zipcodes, in metropolitan counties, and non-white. 
We show that correcting for underreporting is important to accurately quantify these disparities, and that failing to do so yields less plausible estimates. Our method is broadly applicable to underreported conditions in women's health, as well as to gender biases beyond healthcare.  

\end{abstract}

\clearpage

\linespread{1.0}\selectfont
\FloatBarrier
\pagenumbering{gobble}


\begin{figure}[t!]
  \centering
  \includegraphics[width=\columnwidth]{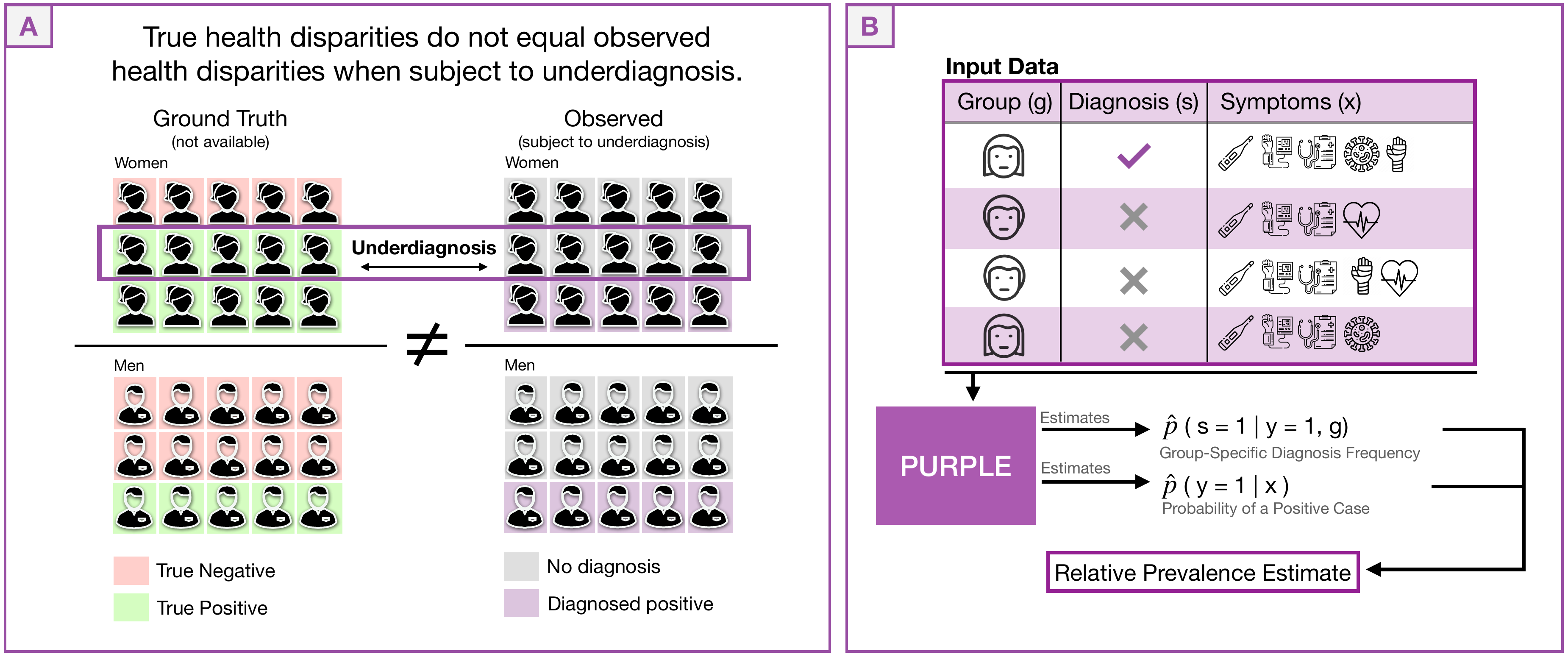}
  \caption{\textbf{Underreporting can skew observed relative prevalences and conceal health disparities. \texttt{PURPLE} is designed to estimate the relative prevalence while correcting for underreporting.} \textbf{A) Underreporting leads to inaccurate observed relative prevalences.} Understanding the relative prevalence of a health condition between groups $g$---for example, men and women---is important to effective medical care. However, these estimates are often based on diagnoses $s$ (i.e. a diagnosed positive or no diagnosis) instead of the true patient state $y$ (sick vs. not sick). Underreporting, which is known to vary by demographic groups, leads to inaccurate relative prevalence estimates that can hide the groups most affected by a condition. \textbf{B) \texttt{PURPLE} uses data on patient diagnoses $s$, symptoms $x$, and group membership $g$ to accurately estimate the relative prevalence of a condition.} \texttt{PURPLE} first estimates the group-specific diagnosis probability, $p(s=1|y=1,g)$, and disease likelihood, $p(y=1|x)$, up to constant multiplicative factors, and then combines these estimates to compute the relative prevalence. We show this is possible under three widely-made assumptions: no false positives, random diagnosis within groups, and constant $p(y=1|x)$ between groups.}
  \label{fig:intro}
\end{figure}


\begin{figure}[t!]
  \centering
  \includegraphics[width=\columnwidth]{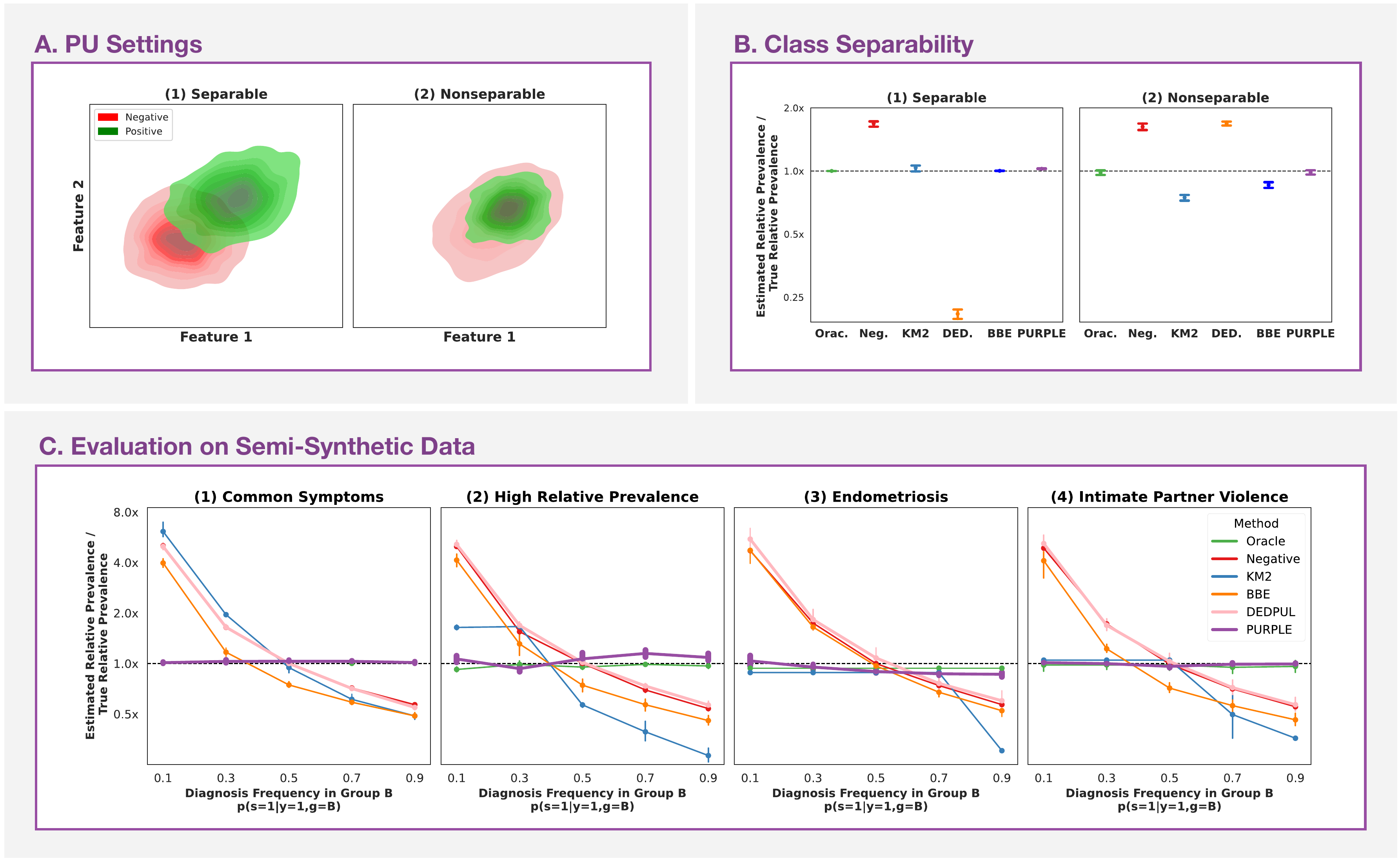}
  \caption{\textbf{Validation of \texttt{PURPLE} on synthetic and semi-synthetic data.} \textbf{A) Methods in positive-unlabeled learning commonly make assumptions about the separability of the positive and negative distributions.} Settings in which underreporting occurs map directly to work in positive-unlabeled learning, in which learning algorithms have access to a set of positive labeled examples and an unlabeled mixture of positive and negative examples. Most works in positive-unlabeled learning assume A (left panel), or a positive subdomain, while no method can accommodate the distributions pictured in the right panel. \texttt{PURPLE} makes no assumptions about the separability of the positive and negative  distributions, and instead assumes that $p(y=1|x)$ remains constant across patient subgroups. \textbf{B) \texttt{PURPLE} accurately recovers the relative prevalence on both separable and nonseparable synthetic data.} The vertical axis plots the ratio of estimated relative prevalence to true relative prevalence, with 1 (dotted line) indicating perfect performance. We report variation across 5 randomized train, validation and test splits. \emph{Negative}, \emph{KM2}, \emph{BBE}, and \emph{DEDPUL} baselines do not always accurately estimate the relative prevalence, especially on nonseparable data. \emph{Oracle} is impossible to implement in practice because it relies on ground truth labels $y$ which are not available; it is provided as a metric for ideal performance. \textbf{C)} \textbf{\texttt{PURPLE} recovers the relative prevalence accurately in simulations based on real health data.} We generate semi-synthetic data by using patient visits from MIMIC-IV\cite{johnson2020mimic} and simulating a disease label given a set of symptoms. This allows us to test \texttt{PURPLE} on a real, high-dimensional distribution of symptoms while retaining access to ground truth labels. Each dataset simulates disease likelihood on the basis of a different symptom set: \textbf{(1)} symptoms that appear most frequently, \textbf{(2)} symptoms which occur frequently in one group but not the other, \textbf{(3)} symptoms that co-occur frequently with endometriosis, and \textbf{(4)} symptoms known to indicate risk of intimate partner violence based on past literature. 
  We define group A to be Black patients, and group B to be white patients.
  Across symptom sets, and a range of group-specific diagnosis frequencies, \texttt{PURPLE} produces more consistently accurate relative prevalence estimates than existing work. Two semisynthetic experiments involving real conditions in women's health---endometriosis and intimate partner violence---demonstrate the potential to apply \texttt{PURPLE} to conditions in women's health  and produce accurate, actionable relative prevalence  estimates.
  }
  \label{fig:separability_cov_shift}
\end{figure}





\begin{figure}[t!]
  \centering
  \includegraphics[width=\columnwidth]{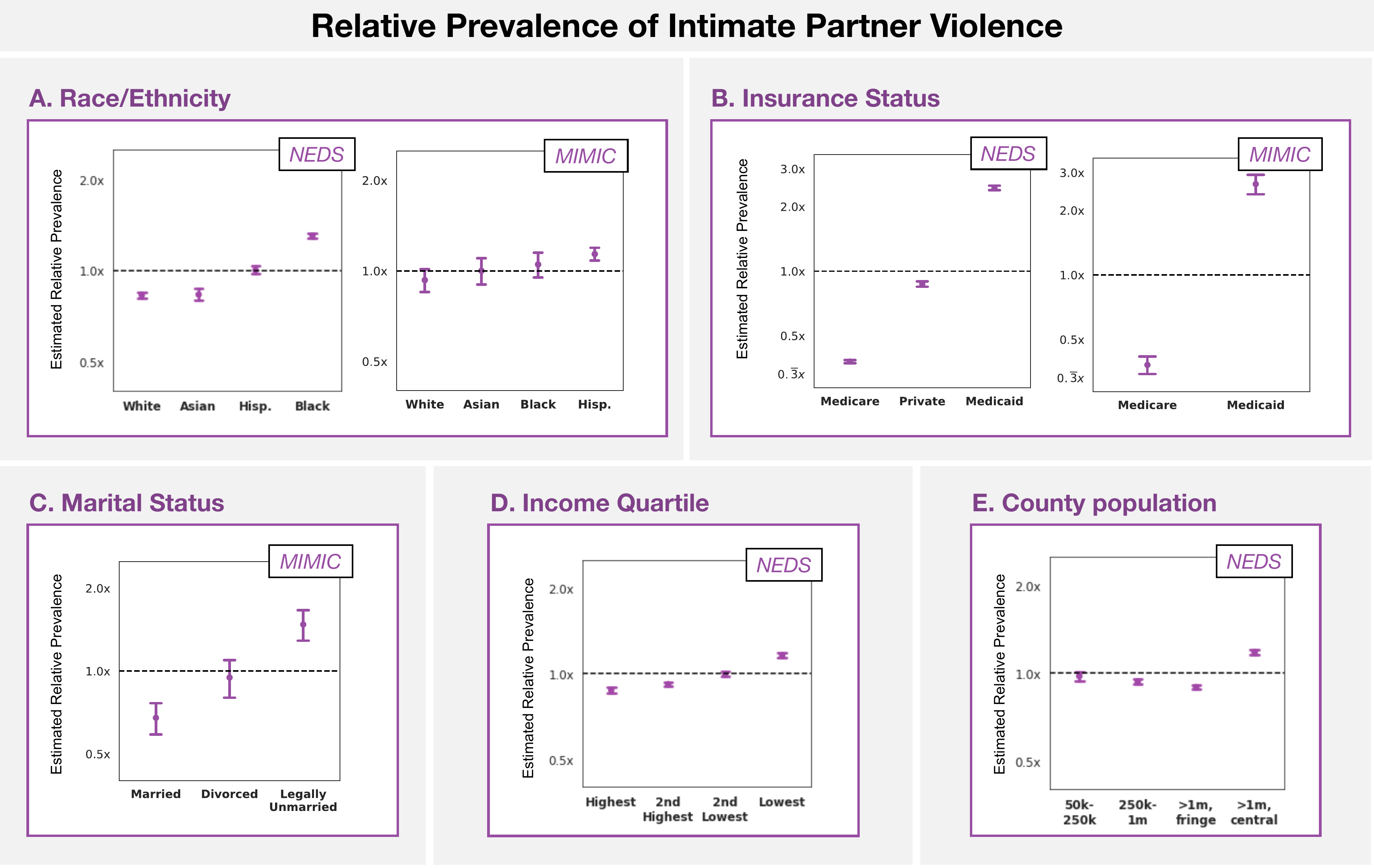}
  \caption{\textbf{Estimates of the relative prevalence of intimate partner violence across demographic subgroups.} We apply \texttt{PURPLE} to two large emergency department datasets: NEDS\cite{NEDS} and MIMIC-IV\cite{johnson2021mimic}. We compute each relative prevalence with respect to each group's complement (i.e. estimating the prevalence of IPV among married patients vs. non-married patients). Relative prevalences are higher among non-white patients overall (A) though these disparities are noisily estimated in the MIMIC dataset and the ordering across race groups is not completely consistent. We also find higher prevalences among patients who are on Medicaid (B), not legally married (C), in lower-income zipcodes (D), and in metropolitan counties (E). Counties with more than 1 million residents are further differentiated by how urban they are (as ``central'' or ``fringe). Error bars report the standard deviations for each estimate across 5 randomized train/test splits of each dataset. Panels C, D, and E are reported on only a single dataset due to different demographic feature availability in the two datasets; note the slightly larger y-axis range on Panel B due to the greater range of the estimates.}
  \label{fig:ipv}
\end{figure}

\FloatBarrier
\clearpage

\pagenumbering{arabic}


\linespread{1.6}\selectfont
\section*{Introduction}

There are enormous disparities in women's health across race, age, socioeconomic status, and other dimensions.
Mitigating these disparities requires accurate estimates of the extent to which  a medical condition disproportionately affects different groups.
The \emph{relative prevalence} does so by capturing how much more frequently a condition occurs in one group compared to another --- $\frac{\text{prevalence in group A}}{\text{prevalence in group B}}$ --- with high relative prevalence estimates suggesting concrete areas to increase funding, research, and resources. Public health decisions often rely on such estimates to develop, allocate, and advocate for interventions. For example, research revealing startling disparities in maternal mortality between Black and white women\cite{macdorman2017trends} led to Congressional policy that has invested billions in funding towards evidence-based interventions to improve Black maternal health\cite{WH}. 


However, it remains challenging to produce accurate relative prevalence estimates for many conditions in women's health due to widespread \emph{underreporting}
 (Fig. \ref{fig:intro}A). With underreported health conditions, only a small percentage of true positives may be labeled as positive; worse, the probability of correctly diagnosing a positive case can vary by group\cite{geiger2003racial}. This is especially relevant to intimate partner violence, a notoriously underreported condition: true cases are only correctly diagnosed an estimated $\sim$25\% of the time, and this probability varies across racial groups\cite{schafer2008using}. Underreporting in one group and not another can mask health disparities, making it appear that a condition is equally prevalent in two populations when it is not, or more prevalent in one population than in another when it is not. These errors obscure where resources are most needed, and consequently inhibit the development of effective health policy.

Efforts in both epidemiology and machine learning have addressed these challenges, but often rely on either data which is unavailable, or assumptions which are unrealistic in women's health contexts (refer to Sec. \ref{sec:related_work} for a detailed discussion of related work). Epidemiological work attempts to quantify true prevalences in the context of imperfect diagnostic tests, and commonly assumes the presence of information which is not always available (for example, ground truth annotations\cite{lyles2011validation,sekar1949method,simeone1995plan,hay2003estimating,mckeganey1992female}, multiple tests\cite{hui1980estimating,walter1988estimation, pepe2007insights},  or informative priors\cite{lewis2012association}). For conditions in women's health, we often have no access to ground truth, only a single diagnosis per patient, and no notion of how accurate that diagnosis truly is. The machine learning literature has modeled underreporting using the positive-unlabeled (PU) learning framework, which assumes that only some positive cases are correctly labeled as positive, and the unlabeled examples consist of both true negatives and unlabeled true positives.  In order to recover prevalence in the presence of underdiagnosis, many PU learning methods assume that there is a region of the feature space where cases are certain to be true positives.  However, this is a restrictive assumption which, while potentially suitable in other PU settings, is unlikely to hold in health data\cite{chen2017machine} because it is rare that a set of symptoms corresponds to a health condition with 100\% certainty. This is especially true in the context of intimate partner violence, where symptoms are frequently not specific to a particular condition (for example, pregnancy complications, which are well-known to occur at higher rates among IPV patients\cite{davidov2015united}, could suggest a number of underlying conditions, rather than just one).

Here, we present \texttt{PURPLE} (\textbf{P}ositive \textbf{U}nlabeled \textbf{R}elative \textbf{P}reva\textbf{L}ence \textbf{E}stimator), a method that is complementary to the epidemiology and PU learning literature. In contrast to epidemiological approaches, it requires no external information (for example, the sensitivity and specificity of a test) to recover the relative prevalence of an underreported disease. In contrast to the PU learning literature, \texttt{PURPLE} relies on no assumptions about the symptom distributions of positive and negative cases. \texttt{PURPLE} is designed to address underreporting in intimate partner violence, and women's health more broadly, by estimating the relative prevalence of the condition given three assumptions: 1) no false positive diagnoses; 2) random diagnosis within group; and 3) constant $p(y=1|x)$ between groups, i.e., that the probability of having a disease conditional on symptoms remains constant across groups. The first two assumptions are standard in PU learning; the third, which is specific to our method, replaces PU assumptions about the separability of the positive and negative classes. We show that if these assumptions are satisfied, it is possible to recover the relative prevalence even if it is not possible to recover the absolute prevalence: that is, $\frac{\text{prevalence in group A}}{\text{prevalence in group B}}$ can be estimated even if neither the numerator nor denominator can. \texttt{PURPLE} does this by jointly estimating the conditional probability that a case is a true positive given a set of symptoms, and the diagnosis probability (i.e., the probability a positive case is diagnosed as such; Fig. \ref{fig:intro}B). We demonstrate via experiments on synthetic and real health data that \texttt{PURPLE} recovers the relative prevalence more accurately than do existing methods. We provide methods for checking whether \texttt{PURPLE}'s underlying assumptions hold, and show that even under a plausible violation of the assumptions \texttt{PURPLE} still provides a useful lower bound on the magnitude of disparities. 

Having validated \texttt{PURPLE}, we use it to estimate the relative prevalence of the condition which motivates this work --- intimate partner violence (IPV) --- in two widely-used datasets of electronic health records, which  together describe millions of emergency department visits: MIMIC-IV\cite{johnson2021mimic} and NEDS\cite{NEDS}. 
Across both datasets, we find higher prevalences of IPV among patients who are on Medicaid than among those on Medicare. Relative prevalences are higher among non-white patients (though these disparities are noisily estimated in the MIMIC dataset). We also quantify the relative prevalence of IPV across income quartiles, marital statuses, and the rural-urban spectrum, finding that IPV is more prevalent among patients who are lower-income, not legally married, and in metropolitan counties. Finally, we show that \texttt{PURPLE}'s corrections for underreporting are important: they yield more plausible estimates of how relative prevalence varies with income than estimation methods which do not correct for underreporting. Specifically, \texttt{PURPLE} estimates that the relative prevalence of IPV decreases with income, consistent with prior work\cite{rennison2000bureau, bonomi2014intimate,abramsky2019women}. In contrast, failing to correct for underdiagnosis (i.e. computing relative prevalence estimates using observed diagnoses) yields estimates which do not show any consistent trend with respect to income, and which are harder to explain. Overall, this analysis contributes to the literature on IPV disparities in several ways: it uses some of the largest and most recent samples; evaluates robustness across multiple datasets; and corrects for underreporting.
 
 
 Together, our analyses illustrate how \texttt{PURPLE} is a general method for estimating relative prevalences in the presence of underreporting, allowing practitioners to discover and quantify group-specific disparities in a wide range of settings in women's health and beyond.

\section*{Results}
\setcounter{section}{2}
\setcounter{subsection}{0}

Here, we introduce \texttt{PURPLE}, a method to quantify disparities in prevalence of a health condition between groups given only positive and unlabeled data. A key idea underpinning our method is that knowing the exact prevalence in a group is not necessary to calculate the \textit{relative} prevalence across groups: one can estimate the fraction $\frac{\text{prevalence in group A}}{\text{prevalence in group B}}$ without knowing its numerator or denominator. We adopt terminology standard in the PU learning literature and assume that we have access to three pieces of data for the $i$th example: a feature vector $x_i$; a group variable $g_i$; and a binary observed label $s_i$. We let $y_i$ denote the true (unobserved) label. In healthcare, example $i$ may correspond to a specific patient and their  presenting symptoms ($x_i$), race ($g_i$), and observed diagnosis ($s_i$). Here, $y_i$ corresponds to whether the patient truly \emph{has} the medical condition. We first introduce \texttt{PURPLE} and then validate our approach on synthetic and semi-synthetic data.

\subsection*{Overview of \texttt{PURPLE}: Positive-Unlabeled Relative Prevalence Estimation}

We provide a conceptual overview of \texttt{PURPLE} here and describe the full details in Sections \ref{sec:relative_prevalence}-\ref{sec:implementation}. 
\texttt{PURPLE} is designed to address how, for underreported women's health conditions, the ratio of \emph{diagnosis rates} between demographic groups may not equal the ratio of \emph{true prevalences}, due to differential underreporting across groups. 
To address this, \texttt{PURPLE} first uses the observed data to learn a model of which symptoms correlate with having the condition; so long this relationship between symptoms and the condition remains constant  across groups, we will be able to estimate the relative prevalence. 
Mathematically, \texttt{PURPLE} first estimates $p(y=1|x)$ up to a constant multiplicative factor; second, it uses this estimate to compute the relative prevalence  between groups. We use $p$ to denote the true probabilities in the underlying data distribution and $\hat p$ to denote \texttt{PURPLE}'s estimates of these probabilities.



\begin{enumerate}
    \item \textbf{Estimate $p(y=1|x)$ up to a constant factor.} \texttt{PURPLE} fits the following model: 
\begin{align}
\underbrace{\hat p(s=1|g,x)}_{\substack{\text{probability patient}\\\text{is diagnosed}}} & = \underbrace{\hat{p}(y=1|x)}_{\substack{\text{probability patient}\\ \text{truly has condition}}} \cdot \underbrace{\hat{p}(s=1|y=1,g)}_{\substack{\text{probability true positives}\\ \text{are correctly diagnosed}}}
 \label{eqn:probabilistic_model}
\end{align}
    In other words, \texttt{PURPLE} models the probability a patient is diagnosed with a condition as the product of two terms: the probability the patient truly has the condition, and the probability that true positives are correctly diagnosed. The first term is constant across groups $g$ while the second can vary, accounting for underdiagnosis. This decomposition is valid under three assumptions which we discuss below. \trackchange{To estimate the two terms on the right hand side of Equation \ref{eqn:probabilistic_model}, we parameterize the first term as a single-layer neural network and the second as a constant $c_g \in [0, 1]$ for each group $g$. We optimize these parameters by minimizing the cross-entropy loss between the predicted $\hat{p}(s=1|g,x)$ and the true $p(s=1|g,x)$, which is possible because $s$, $g$, and $x$ are all observed (Sec. \ref{sec:implementation})}. Note that we can only estimate both terms on the right hand side up to a constant multiplicative factor, because multiplying the first term by a non-negative $\beta$ and dividing the second term by $\beta$ leaves $\hat p(s=1|g,x)$ unchanged. 
    \item \textbf{Estimate the relative prevalence using $\hat{p}(y=1|x)$.} Fortunately, even though $\hat{p}(y=1|x)$ is only correct up to a constant multiplicative factor, this suffices to estimate the relative prevalence $\frac{p(y=1|g=a)}{p(y=1|g=b)}$, as we derive in Sec. \ref{sec:relative_prevalence}. Specifically, our estimator of the relative prevalence is
    \begin{equation}
        \label{eqn:rp_estimate}
        \frac{\sum_x \hat{p}(y=1|x)\hat{p}(x|g = a)}{\sum_x \hat{p}(y=1|x)\hat{p}(x|g = b)}
    \end{equation}
    In practice, this is simply the mean value of $\hat{p}(y=1|x)$ for samples from group $a$ divided by the mean value of $\hat{p}(y=1|x)$ for samples from group $b$.
\end{enumerate}

As discussed above, it is impossible to make progress in PU settings without assumptions\cite{bekker2020learning}. Our estimation procedure relies on three assumptions: 1) observed positives are true positives (the \emph{positive-unlabeled} assumption common to all PU methods), 2) within each group, diagnosis $s$ depends only on $y$ (the \emph{random diagnosis within group} assumption, commonly made in PU settings) and 3) the probability of having a disease conditional on symptoms remains constant across groups (the \emph{constant $p(y|x)$} assumption, common to work in both domain adaptation\cite{sugiyama2007covariate,quinonero2009dataset} and healthcare\cite{nestor2019feature}). Details about the required assumptions can be found in Sec.  \ref{sec:assumptions}. We also provide checks to assess whether the assumptions hold (Sec. \ref{sec:assumption_checks}), and show that even under plausible violations of these assumptions, \texttt{PURPLE} is guaranteed to produce a lower bound on the true magnitude of disparities (Sec. \ref{sec:cov_shift_assumption}). \trackchange{An illustration of \texttt{PURPLE}'s behavior under violations of the PU assumption and random-diagnosis-within-group assumption is available in Sec. \ref{sec:pu_violations} and Sec. \ref{sec:scar_violations}, respectively}. We provide the full derivation of our estimation procedure in Sec. \ref{sec:methods}.




\subsection*{\texttt{PURPLE} recovers the true relative prevalence on synthetic data}

Prior to applying \texttt{PURPLE} to estimate the relative prevalence of IPV, we confirm that the method can correctly recover the true prevalence on synthetic data where the true relative prevalence is known, a standard machine learning check. We compare \texttt{PURPLE} to four previous machine learning methods (Sec. \ref{sec:baselines}) drawn from the literature on PU learning, where estimating prevalence is a critical step\cite{bekker2019beyond}. We generate the synthetic data by simulating group-specific features ($p(x|g)$) , and labels using a decision rule ($p(y|x)$). The two groups, $a$ and $b$, correspond to 5D Gaussian distributions with different means (see Sec. \ref{sec:synth_data_generation} for full data generation details). 




 Figure \ref{fig:separability_cov_shift}B compares \texttt{PURPLE}'s performance to performance of the other methods on purely synthetic data. We evaluate each approach in both separable (in which the datapoints with $y=1$ and the datapoints with $y=0$ can be perfectly separated in the feature space $x$) and non-separable settings. We perform this comparison because existing methods rely on separability assumptions which often do not hold in realistic health settings\cite{chen2017machine}. 
 \texttt{PURPLE} is the only method that accurately recovers the relative prevalence in both the separable and non-separable settings. We also show that \texttt{PURPLE} maintains consistent performance regardless of  the extent to which $p(x)$, or the distribution of symptoms, differs between groups (Figure \ref{fig:supp_fig_cov_shift}).

\subsection*{\texttt{PURPLE} recovers the true relative prevalence in realistic semi-synthetic health data}

Having established that \texttt{PURPLE} outperforms previous work on synthetic data, we investigate its performance on more realistic data: specifically, MIMIC-IV\cite{johnson2020mimic}, a dataset of electronic health records that describes $\sim$450,000 patient hospital visits between 2008-2018.  
We use these records to generate realistic semi-synthetic data to examine \texttt{PURPLE}'s performance on the high-dimensional, sparse data common in clinical settings. Specifically, we use the patient symptoms $x$ --- encoded as a binary one-hot vector of ICD codes --- to simulate whether the patient truly has the medical condition, $y$. Using data in which we know $y$ allows us to assess how accurately \texttt{PURPLE} recovers the relative prevalence; in contrast, if we did not simulate $y$, we would not have access to ground truth, and could not assess relative prevalence estimates. We simulate $y$ for four settings: 1) a condition with common symptoms, 2) a condition that is less common among Black patients, 3) endometriosis, and 4) intimate partner violence (see Sec. \ref{sec:semi_synth} for full details). 

Across the semi-synthetic settings we consider, the estimation error of previous methods is large, with some methods producing relative prevalence estimates more than 4x the true value (Fig. \ref{fig:separability_cov_shift}C). Further, each previous method produces both overestimates and underestimates of the true relative prevalence depending on how underreported the medical condition is. In contrast, \texttt{PURPLE} remains accurate across the different settings.



\section*{Quantifying the relative prevalence of intimate partner violence}
\setcounter{section}{4}
\setcounter{subsection}{0}

We have validated \texttt{PURPLE}'s accuracy in recovering the relative prevalence by using synthetic and semi-synthetic datasets where the true relative prevalence is known. We now use \texttt{PURPLE} to estimate relative prevalence on two real datasets where the true relative prevalence is \emph{unknown}. Specifically, we apply \texttt{PURPLE} to quantify the relative prevalence of the underdiagnosed condition motivating this work --- intimate partner violence (IPV) --- across different demographic groups.


\paragraph{Datasets} We conduct our study using two widely-used datasets of emergency department visits: MIMIC-IV ED\cite{johnson2021mimic} and the 2019 Nationwide Emergency Department Sample (NEDS)\cite{NEDS}. MIMIC-IV describes 293,297 emergency department visits to a single, Boston-area hospital; NEDS is a nationwide sample which is approximately one hundred times as large (it contains 33.1 million emergency department visits, which, when reweighted, represent the universe of 143 million US emergency department visits in 2019). We assess results across multiple datasets to verify the robustness of the disparities we observe. Because our sample consists of emergency department visits, we estimate the relative prevalence of IPV \emph{conditional on going to the emergency department} --- in particular, our data does not allow us to quantify disparities among populations who do not interact with the healthcare system at all\cite{riley2012health}. Relative prevalence estimates among patients who visit emergency departments, however, remain of interest to IPV researchers due to the unique role emergency departments play as a point of care \emph{and} intervention on patients who suffer from IPV\cite{alessandrino2020intimate}.
For both datasets, we filter for female patients because the symptoms associated with IPV in male patients are less well understood and the constant $p(y=1|x)$ assumption may not hold\cite{houry2008differences}; we also filter out patients younger than 18 years old because symptoms that indicate intimate partner violence could be instances of child abuse in this patient subgroup\cite{louwers2011detection,loder2020demographics}. We describe all preprocessing steps in Section \ref{sec:methods-datasets}. All point estimates and uncertainties reported below represent the mean and standard deviation, respectively, across five randomized train/test splits of each dataset. 


\paragraph{Analysis} Results are plotted in Figure \ref{fig:ipv}. (We also verify that \texttt{PURPLE} passes the assumption checks detailed in Section \ref{sec:assumption_checks} in SI Figures \ref{fig:auc_auprc_check}, \ref{fig:calibration_check_mimic}, \ref{fig:calibration_check_neds}). We find, in both datasets, that intimate partner violence is more common among patients on Medicaid compared to patients on Medicare (NEDS relative prevalence $2.44 \pm 0.07$ in Medicaid patients versus $0.37 \pm 0.01$ in Medicare patients; MIMIC-IV relative prevalence $2.65 \pm 0.31$ in Medicaid patients versus $0.38 \pm 0.04$ in Medicare patients). Of course, Medicaid is likely not the \emph{causal} factor underlying IPV risk; rather, it acts as a proxy which identifies populations who are disproportionately affected by IPV. 

Examining racial differences reveals disparities which are smaller and less consistent than disparities by insurance status. In both datasets, white patients have the lowest relative prevalence of the four race groups, and in the NEDS dataset white patients have significantly lower prevalence than non-white patients overall (relative prevalence for white patients vs. non-white patients $0.82 \pm 0.02$). However, in MIMIC-IV, racial disparities are more noisily estimated due to the smaller size of the dataset, yielding an ordering of race groups which is similar but not completely consistent across the two datasets. This attests to the importance of using large samples and assessing results across multiple datasets.


The MIMIC-IV dataset provides information on patient marital status, allowing us to estimate that IPV is more common among patients who are ``Legally Unmarried", who are not officially married but may still be in relationships (relative prevalence $1.48\pm 0.21$). The NEDS dataset provides information on the population density and estimated median household income of areas where patients live. We estimate higher rates of IPV among patients living in central metropolitan counties with population $>$1 million (relative prevalence $1.18\pm 0.02$). We also find that IPV prevalence decreases with income (relative prevalence $1.16 \pm 0.02$ in the bottom income quartile versus $0.87 \pm 0.03$ in the top income quartile).

In Figure \ref{fig:observed_rps}, we report the prevalence of observed IPV diagnoses --- i.e., $p(s=1|g)$ --- without correcting for underdiagnosis. While often the trends are qualitatively similar, in some cases correcting for underdiagnosis is important to yield plausible trends. For example, failing to correct for underdiagnosis produces an inconsistent relationship between IPV prevalence and income which is difficult to reconcile with past work consistently documenting that IPV prevalence decreases with income\cite{rennison2000bureau, bonomi2014intimate,abramsky2019women}. This suggests the importance of using methods, like \texttt{PURPLE}, which attempt to correct for underdiagnosis. Our income results also suggest that IPV is less likely to be correctly diagnosed in lower-income women, a finding that reflects the broader phenomenon of underdiagnosis among lower-income patients, as has been shown in the context of dementia\cite{amjad2018underdiagnosis,bradford2009missed}, asthma\cite{aaron2018underdiagnosis, quinn2006racial}, and depression\cite{swant2007200,lao2016underdiagnosis,sorkin2011underdiagnosed}.

\section*{Discussion} 

In this work, we provide a method for estimating relative prevalence even in the presence of underreporting, a difficult but essential task in healthcare and public health more broadly. We show that we can estimate the relative prevalence even in settings where absolute prevalence estimation is impossible, by exchanging the restrictive separability assumptions typical in the PU learning literature for the constant $p(y=1|x)$ assumption, which is arguably more appropriate in clinical settings. Although this assumption may not hold for all settings---for example, the conditional probability of intimate partner violence is known to be dependent on a patient's age group\cite{pathak2019experience}---it is realistic in many settings, and we provide methods for checking its validity and a lower-bound guarantee even when it fails to hold. Based on these assumptions, we present a method for relative prevalence estimation, \texttt{PURPLE}, a complementary approach to those in the epidemiology and PU learning literature: it works when one does not have the external information which epidemiological methods generally require, and cannot make the separability assumptions PU learning methods rely on. We show \texttt{PURPLE} outperforms previous methods in terms of its ability to recover the relative prevalence on both synthetic and real health data. 

We apply \texttt{PURPLE} to estimate the relative prevalence of intimate partner violence in two widely-used, large-scale datasets of emergency department visits. We find that IPV is more prevalent among patients who are on Medicaid, non-white, not legally married, in lower-income zipcodes, and in metropolitan counties. We also show that correcting for underdiagnosis produces estimates of IPV prevalence across income groups which are more plausible in light of prior work\cite{rennison2000bureau, bonomi2014intimate,abramsky2019women}, highlighting the importance of modeling underdiagnosis. In general, past work on IPV disparities corroborates the plausibility of our findings. Our finding that intimate partner violence is more common among patients on Medicaid compared to patients on Medicare is consistent with earlier results that show that IPV is less common among elderly patients\cite{evans2017diagnosis,gerino2018intimate,pathak2019experience}, and more common among patients who live below the poverty line\cite{cunradi2000neighborhood,bonomi2014intimate,mariscal2020changes}. Past work documenting higher IPV prevalences among unmarried women\cite{wong2016comparison,abramsky2011factors,capaldi2012systematic} and in metropolitan areas\cite{rennison2000bureau,ravi2022survivors,dubois2019intimate} also corroborates the plausibility of our findings. Our finding that IPV is more common among non-white patients is corroborated by some past work\cite{lipsky2009racial,schafer2008using,rennison2000bureau}. However, the fact that we find that racial disparities are smaller and not completely consistent across datasets is also concordant with past work documenting inconsistent racial differences across samples\cite{cho2012racial,mariscal2020changes,nyc2020report,hart2013practical}. This suggests the importance of using large samples, and multiple datasets, to assess how consistently and robustly racial disparities emerge. Overall, our analysis contributes to the literature on IPV disparities by using large samples; evaluating robustness across multiple datasets; and correcting for underreporting.

Our work is motivated by widespread underreporting in women's health, and we foresee numerous opportunities for future work. \texttt{PURPLE} could be applied to obtain relative prevalence estimates for many other health conditions that are known to be underreported, including polycystic ovarian syndrome\cite{hillman2018polycystic}, endometriosis\cite{agarwal2019clinical}, and traumatic brain injuries\cite{prince2017evaluation}. Additionally, quantifying relative prevalence in the presence of underreporting is a problem of interest in many domains beyond healthcare and public health: for example, quantifying the relative prevalence of underreported police misconduct across precincts, or quantifying the relative prevalence of underreported hate speech across demographic groups. We believe that \texttt{PURPLE} can also yield useful insight into disparities in these non-healthcare settings.

\paragraph{Data availability:} Anonymized imaging and clinical data to reproduce results of this study are available online. MIMIC-IV is a publicly available database of emergency department and hospital admissions occuring between 2008 and 2019 at the Beth Israel Deaconness Medical Center. NEDS 2019 is also a publicly available Both datasets are publicly available. MIMIC-IV can be found at: https://physionet.org/content/mimiciv/2.2/. The HCUP NEDS 2019 database is also publicly available at: https://www.hcup-us.ahrq.gov/nedsoverview.jsp. Code to preprocess all datasets and reproduce all experiments can be found at  \href{https://github.com/epierson9/invisible-conditions}{https://github.com/epierson9/invisible-conditions}.

\paragraph{Acknowledgments:} We thank Microsoft Research seminar participants, Irene Chen, Serina Chang, Jacquelyn Campbell, Karthik Chetty, Nikhil Garg, John Guttag, Pang Wei Koh, Allison Koenecke, Nat Roth, Priya Shanmugam, and Harini Suresh for helpful conversations. EP was supported by a Google Research Scholar award, an NSF CAREER award, a CIFAR Azrieli Global scholarship, a LinkedIn Research Award, and a Future Fund Regrant.

\paragraph{Competing interests:} The authors declare no competing financial or non-financial interests.

\paragraph{Author contributions:} E.P. supervised the project. E.P. and D.S. conceived of the presented method, experiments, theory, and verified the results. E.P. and D.S. wrote the manuscript. D.S. conducted the synthetic and semi-synthetic experiments, carried out the case studies for intimate partner violence and content moderation, and created the figures. K.H. conducted analyses of the MIMIC-IV data and the NEDS data.

\clearpage

\nolinenumbers 
\section*{References}
\linespread{1.0}\selectfont
\bibliographystyle{naturemag}
\bibliography{main}
\clearpage

\linenumbers
\linespread{1.6}\selectfont
\renewcommand\thesection{M\arabic{section}}
\renewcommand\thefigure{M\arabic{figure}}
\renewcommand\thetable{M\arabic{table}}
\renewcommand{\figurename}{Figure}
\renewcommand{\tablename}{Table}

\section*{Methods}
\setcounter{section}{0}
\setcounter{subsection}{0}

\section{\texttt{PURPLE}: Positive Unlabeled PrevaLence Estimator}
\label{sec:methods}

\texttt{PURPLE} estimates $p(y=1|x)$ up to a constant multiplicative factor in order to estimate the relative prevalence of a condition $y$. Underlying  this procedure are two insights: first, that estimating $p(y=1|x)$ up to a constant factor suffices to estimate the relative prevalence, and second, that it is possible to produce this estimate using the observed labels $s$, symptoms $x$, and group statuses $g$. We first describe the three assumptions underlying \texttt{PURPLE}, and show how these statements follow from them in Sections \ref{sec:relative_prevalence} and \ref{sec:estimating_likelihood_fn}. We describe implementation details in Sec. \ref{sec:implementation}. We provide checks to determine whether \texttt{PURPLE}'s assumptions hold true (Sec. \ref{sec:assumption_checks}) and show that even under a plausible violation of our assumptions, \texttt{PURPLE} produces a lower bound on the true magnitude of disparities (Sec. \ref{sec:cov_shift_assumption}). 

\subsection{Assumptions}\label{sec:assumptions}

Neither the exact prevalence nor the relative prevalence can be recovered without making assumptions about the data generating process: intuitively, without further assumptions, it is impossible to distinguish between whether a medical condition is truly rare or merely rarely \emph{diagnosed}. We adopt terminology standard in the PU learning literature and assume that we have access to three pieces of data for the $i$th example: a feature vector $x_i$; a group variable $g_i$; and a binary observed label $s_i$. We let $y_i$ denote the true (unobserved) label. In healthcare, example $i$ may correspond to a specific patient and their  presenting symptoms ($x_i$), race ($g_i$), and observed diagnosis ($s_i$). Here, $y_i$ corresponds to whether the patient truly \emph{has} the medical condition. This is an unobserved binary variable and because the medical condition is underreported, not all patients who truly have the condition are diagnosed with it, so $p(s_i=1|y_i=1) < 1$. Because we are interested in health disparities, we focus on groups $g$ defined by sensitive attributes (e.g., gender, race, or socioeconomic status) but our method is applicable to any set of groups for which our assumptions hold. We make three assumptions:


\begin{enumerate}
    \itemsep 0em
    \item \emph{No False Positives}: We assume that examples labeled as positive ($s=1$) are truly positive ($y=1$): i.e., $p(y=0|s=1) = 0$ (and thus, by Bayes' rule, $p(s=1|y=0) = 0$). This is the positive unlabeled assumption and is the foundational assumption of PU learning methods\cite{elkan2008learning}.
    \item \emph{Random Diagnosis within Groups}: We assume that positive examples within a specific group are equally likely to be labeled as positive: $p(s=1|y=1, g=a) = c_a$, where $c_a$ represents the diagnosis frequency of group $a$ (i.e. the probability that a positive case is diagnosed as such). Borrowing terminology from PU learning,  this amounts to the commonly made Selected-Completely-at-Random assumption\cite{bekker2018estimating} within each group. We allow $c_a$ to vary across groups to allow for group-specific underdiagnosis rates.
    \item \emph{Constant $p(y=1|x)$ between Groups}: We assume that $p(y=1|x)$ remains constant across groups: examples in different groups with the same features are equally likely to be true positives. In the medical setting, this means that patients in different groups with the same symptoms have the same probability of truly having a condition. This is equivalent to assuming only \emph{covariate shift} between groups, a commonly made assumption in the literature on domain adaptation\cite{sugiyama2007covariate,quinonero2009dataset} and healthcare\cite{nestor2019feature}.
\end{enumerate}

Notably, we make no assumptions about the separability of the positive and negative distributions. Past work in PU learning has shown that the true prevalence $p(y=1)$ can be recovered under a restrictive set of assumptions about the structure of the positive and negative distributions\cite{jain2016estimating}, which we refer to as ``separability assumptions". Many PU learning methods assume that the positive distribution is not completely contained within the negative distribution: in healthcare, this means there is a region of the feature space where all examples are true positives\cite{elkan2008learning,du2014class, northcutt2017learning,ramaswamy2016mixture} (Fig. \ref{fig:separability_cov_shift}A). This assumption is unrealistic in medical settings because it is unlikely that a set of symptoms maps to a diagnosis with 100\% probability\cite{chen2021probabilistic}, and as a result, \texttt{PURPLE} makes no such assumption. 

While it is necessary to make assumptions to infer the relative prevalence, no assumptions will hold on all datasets, a point we consider in the Discussion. To ensure \texttt{PURPLE} is applied to appropriate datasets, we provide two checks for violations of \texttt{PURPLE}'s assumptions (Sec. \ref{sec:assumption_checks}). We also show that even under a plausible violation of the \emph{Constant $p(y=1|x)$} assumption, \texttt{PURPLE} provides a useful \emph{lower bound} on the magnitude of health disparities (Sec. \ref{sec:cov_shift_assumption}).

\subsection{Deriving the Relative Prevalence}
\label{sec:relative_prevalence}

Here we show that a  constant factor multiplicative approximation of $p(y=1|x)$ recovers the relative prevalence between groups $a$ and $b$ ($\rho_{a,b}$) exactly. The derivation is as follows:

\begin{small}
\begin{eqnarray}
    \rho_{a,b} &\coloneqq& \frac{p(y = 1 | g = a)}{p(y = 1 | g = b)} \label{eqn:first_ratio}\\
    &=& \frac{ \sum_x p(y=1|x, g=a)p(x|g = a)}{ \sum_x p(y=1|x, g=b)p(x|g = b)} \label{eqn:ratio_expanded}\\
    &=& \frac{ \sum_x p(y=1|x)p(x|g = a)}{\sum_x p(y=1|x)p(x|g = b)} \label{eqn:cov_shift}\\
    &=& \frac{\sum_x \hat{p}(y=1|x)p(x|g = a)}{\sum_x \hat{p}(y=1|x)p(x|g = b)} \label{eqn:p_estimate}\\
    && \notag \text{for all } \hat{p}(y=1|x) \propto p(y=1|x) \
\end{eqnarray}
\end{small}
where Eqn. \ref{eqn:cov_shift} follows from the constant $p(y=1|x)$ assumption and Eqn. \ref{eqn:p_estimate} follows because estimates of $p(y=1| x)$ up to a constant multiplicative factor will yield a constant term in the numerator and denominator which cancels. Thus, estimates of $p(y=1|x)$ up to a constant multiplicative factor suffice to compute the relative prevalence. $p(x|g)$ is directly observable from the data, so we can  estimate the numerator as the mean of $\hat{p}(y=1|x)$ over all $x$ in group $a$, and similarly estimate the denominator as the mean of $\hat{p}(y=1|x)$ over all $x$ in group $b$. 

\subsection{Estimating $p(y=1|x)$ up to a constant multiplicative factor}
\label{sec:estimating_likelihood_fn}

We have shown that, if we can estimate $p(y=1|x)$ up to a constant multiplicative factor, we can use this estimate to compute the relative prevalence $\rho_{a,b}$. Now we show how to estimate $p(y=1|x)$ up to a constant multiplicative factor. We do so by applying our three assumptions to derive an decomposition for $p(s=1|x,g)$: 

\begin{small}
\begin{align}
    p(s=1|x,g) =& \quad p(y=1|x,g)p(s=1|y=1,x,g) \label{eqn:estimate_pu_expansion}  \\  \notag +& \quad p(y=0|x,g)p(s=1|y=0,x,g) \\
    =& \quad p(y=1|x,g)p(s=1|y=1,x,g) \label{eqn:estimate_pu_application} \\
    =& \quad p(y=1|x,g)p(s=1|y=1,g) \label{eqn:estimate_random_labeling}\\
    =& \quad p(y=1|x)p(s=1|y=1,g) \label{eqn:estimate_cov_shift} 
\end{align}
\end{small}
Applying the \emph{No False Positives} assumption allows us to remove the second term in Eqn. \ref{eqn:estimate_pu_expansion}, producing Eqn. \ref{eqn:estimate_pu_application}. The \emph{Random Diagnosis within Groups} assumption removes the dependence of the diagnosis probability on $x$, leading to Eqn. \ref{eqn:estimate_random_labeling}. The \emph{Constant $p(y=1|x)$} assumption leads to Eqn. \ref{eqn:estimate_cov_shift}. 

Thus, $p(s=1|x,g)$ can be decomposed as the product of two terms: the probability the patient truly has the condition given their symptoms, $p(y=1|x)$, and the probability that true positives are correctly diagnosed, $p(s=1|y=1,g)$. The fact that the second term varies across groups accounts for group-specific underdiagnosis. This decomposition can be fit via maximum likelihood estimation with respect to the empirical $p(s=1|x, g)$, since $s$, $x$, and $g$ are observed. Note that this only allows estimation of the two terms on the right side of Eqn. \ref{eqn:estimate_cov_shift} up to constant multiplicative factors, since we can multiply $p(y=1|x)$ by a non-negative $\beta$ and divide $p(s=1|y=1, g)$ by $\beta$ while leaving our estimate of $p(s=1|x, g)$ unchanged. However, constant-factor estimation of $p(y=1|x)$ suffices to estimate the relative prevalence. Concretely, we estimate  $p(y=1|x)$ and $p(s=1|y=1,g)$ up to constant multiplicative factors by fitting to $p(s=1|x,g)$; we then use our constant-factor estimate of $p(y=1|x)$ to estimate the relative prevalence as described in \S \ref{sec:relative_prevalence}. 
 
We note that the probabilistic model described by  Eqn. \ref{eqn:estimate_cov_shift} has been previously applied to estimate absolute prevalence in PU settings\cite{bekker2019beyond}. Our novel contribution is to derive a precise set of assumptions in which this probabilistic model can be used to estimate \emph{relative} prevalence, and provide an estimation method to do so. 

\subsection{Implementation}
\label{sec:implementation}

Thus far, we have shown that it is possible to estimate the relative prevalence of an underreported condition by estimating $p(y=1|x)$ up to a constant factor and provided a way to conduct this estimation given only the observed data. One can apply \texttt{PURPLE} to a new dataset in two steps:

\begin{enumerate}
    \item Estimate $p(y=1|x)$ up to a constant multiplicative factor using the observed diagnoses, and the following probabilistic model:
    \begin{equation}
        \hat{p}(s=1|g,x) = \hat{p}(y=1|x)\hat{p}(s=1|y=1,g)
    \end{equation}
    \item Plug our constant multiplicative factor estimate, $\hat{p}(y=1|x)$, into Eqn. \ref{eqn:p_estimate} to produce the relative prevalence estimate. Specifically, we estimate the relative prevalence $\rho_{a, b}$ as:
    \begin{equation}
       \frac{\sum_x \hat{p}(y=1|x)\hat p(x|g = a)}{\sum_x \hat{p}(y=1|x)\hat p(x|g = b)}
    \end{equation}
    In practice, we can compute this fraction simply by taking the mean value of $\hat{p}(y=1|x)$ in each group to compute the numerator and denominator.
\end{enumerate}

We implement the model in PyTorch\cite{paszke2017automatic} using a single-layer neural network to represent $\hat p(y=1|x)$ and group-specific parameters $c_{g}=\hat p(s=1|y=1,g)$ for each group $g$. \trackchange{Note that a singe layer neural network, followed by a logistic activation, is functionally equivalent to a logistic regression, as they both learn a linear transformation of the input features followed by a logistic transformation to produce a predicted probability of the positive class.} We train the model using the Adam optimizer with default parameters (i.e. a learning rate of .001, epsilon of $10^{-8}$, and weight decay of 0) and implement early stopping based on the cross-entropy loss on the held-out validation set. For the semi-synthetic and real data, we use L1 regularization because these experiments are conducted on high-dimensional vectors, most of which we expect to be unrelated to the medical condition, and select the regularization parameter $\lambda \in$ [$10^{-2}$, $10^{-3}$, $10^{-4}$, $10^{-5}$, $10^{-6}$, $0$] using the held-out validation set by maximizing the AUC with respect to the diagnosis labels $s$. While we use a single-layer neural network because our symptoms $x$ are one-hot encoded and we do not anticipate interactions between symptoms, our approach is general and could be applied with deeper neural network architectures to accommodate interactions and nonlinearities. 

\subsection{Assumption Checks}
\label{sec:assumption_checks}

Like all PU learning methods, \texttt{PURPLE} must rely on assumptions. To prevent users from applying \texttt{PURPLE} to datasets where these assumptions do not hold, we provide two empirical tests whose failure implies at least one of the underlying assumptions fails:

\begin{itemize}
    \item \emph{Compare model fit of \texttt{PURPLE} to  unconstrained model.} If \texttt{PURPLE}'s assumptions hold, the diagnosis likelihood $p(s=1|x,g)$ decomposes as the product of two terms: $p(s=1|x,g)=p(y=1|x)p(s=1|y=1,g)$. This is a constrained model of $p(s=1|x,g)$: for example, it does not allow for interaction terms between group $g$ and symptoms $x$. We can compare the performance of \texttt{PURPLE} to a fully unconstrained model for $p(s=1|x,g)$ which allows, for example, these interaction terms. If the unconstrained model better fits the data, metrics including the AUC and AUPRC will be higher on a held-out set of patients. If the constrained and constrained models exhibit similar performance, it is still possible for one of the assumptions to not be true; however, if the models exhibit different performance, it is a sign that \texttt{PURPLE}'s assumptions are not appropriate.
    \item \emph{Compare calibration across groups}. \texttt{PURPLE} estimates a probabilistic model of diagnosis, $p(s=1|x,g)$, which means we can check how well the outputted probabilities reflect the real data by examining model calibration, a standard check\cite{steyerberg2010assessing}. Concretely, we expect that a proportion $z$ of examples that our model gives a probability $z$ of receiving a positive diagnosis truly receive a positive diagnosis, and we expect this to be true for each group. Violations of \texttt{PURPLE}'s assumptions will often cause group-specific calibration to not hold. For example, if $p(y=1|x)$ differs between groups beyond a scalar constant factor, \texttt{PURPLE}'s estimate of $p(s=1|x, g)$ cannot be correct for both groups (since \texttt{PURPLE} assumes $p(y=1|x)$ remains constant). 
\end{itemize}

We note that these assumption checks cannot rule out all forms of model misspecification---and, indeed, no assumption checks can. Since only $x, g$, and $s$ are observed, it is impossible to prove anything about the distribution of $y$. However, the assumption checks will rule out some forms of model misspecification, and guide users away from datasets where applying $\texttt{PURPLE}$ is clearly inappropriate. 

\subsection{Robustness to Violations of the Constant $p(y|x)$ Assumption}
\label{sec:cov_shift_assumption}

In this section we show that under a plausible violation of the central new assumption of our work (constant $p(y=1|x)$ between groups), \texttt{PURPLE} produces a \emph{lower bound} on the magnitude of disparities. This lower bound is useful because we can be confident that if \texttt{PURPLE} infers that a group suffers disproportionately from a condition, that is in fact the case, and we can be confident in targeting policy to that group.

Specifically, we relax the assumption of constant $p(y=1|x)$ across groups by assuming that if group $A$ has a higher overall prevalence of a condition than group $B$ --- i.e., $p(y=1|g=A) > p(y=1|g=B)$ --- group $A$ also has a higher prevalence by a constant factor given the same set of symptoms --- i.e., $p(y=1|x,g = A) = \alpha \cdot p(y=1|x, g=B), \alpha > 1$. This assumption is a reasonable one: when a condition is more prevalent in one group than another, the same symptoms plausibly correspond to higher posterior probabilities $p(y=1|x)$ in the disproportionately affected group. For example, female patients are more likely than male patients to be victims of intimate partner violence overall\cite{houry2008differences}, and if a woman and a man arrive in a hospital with the same injuries, doctors are plausibly more likely to suspect intimate partner violence as the cause of the woman's injuries.


\paragraph{Proof.} Without loss of generality, assume that group $A$ is the group with higher overall disease prevalence: $p(y=1|g=A) > p(y=1|g=B)$. We make the following assumptions: 

\begin{enumerate}
\itemsep 0em
\item Disease prevalence in group $A$ conditional on symptoms is higher by a constant multiplicative factor: $p(y=1|x,g=A) = \alpha \cdot p(y=1|x,g=B), \alpha > 1$.
\item The other \texttt{PURPLE} assumptions hold: that is, $p(y=1|s=1) = 1$ (PU assumption) and $p(s=1|y=1,g,x) = p(s=1|y=1,g)$ (random diagnosis within groups). 
\end{enumerate}
Under these assumptions, we show that \texttt{PURPLE}'s estimate provides a lower bound on the true relative prevalence. As before, we use $p$ to denote the true probabilities in the underlying data distribution and $\hat p$ to denote \texttt{PURPLE}'s estimates of these probabilities. We have
\begin{eqnarray}
    p(s=1|x, g=A) &=& p(s=1|y=1, g=A) \cdot p(y=1|x, g=B) \cdot \alpha \nonumber \\
    p(s=1|x, g=B) &=& p(s=1|y=1, g=B) \cdot p(y=1|x, g=B)\nonumber
\end{eqnarray}
\texttt{PURPLE} estimates $\hat p(s=1|x, g) = \hat p(s=1|y=1, g) \cdot \hat p(y=1|x)$. It can minimize the cross-entropy loss by achieving an estimate $\hat p(s=1|x, g)$ which matches the true $p(s=1|x, g)$ by setting:
\begin{eqnarray}
    \hat p(y=1|x) &=& \beta \cdot p(y=1|x, g=B) \nonumber \\
    \hat p(s=1|y=1, g=A) &=& \frac{1}{\beta} \cdot \alpha \cdot p(s=1|y=1, g=A) \nonumber \\ 
    \hat p(s=1|y=1, g=B) &=& \frac{1}{\beta} \cdot p(s=1|y=1, g=B) \nonumber
\end{eqnarray}
where $\beta$ is a positive constant (this captures the fact that, as discussed previously, \texttt{PURPLE} only ever estimates $p(y=1|x)$ up to a constant multiplicative factor). In other words, \texttt{PURPLE} can perfectly match the true probabilities $p(s=1|g, x)$ by pushing the variation across groups in $p(y=1|x)$ into $\hat p(s=1|y=1, g)$ (since \texttt{PURPLE} assumes $\hat p(y=1|x)$ remains constant across groups.) Given its estimate $\hat p(y=1|x)$, \texttt{PURPLE}'s estimate $\hat {\rho}_{A,B}$ of the relative prevalence is
\begin{eqnarray}
\hat {\rho}_{A,B} &=& \frac{\sum_{x \in \mathcal{A}} \hat{p}(y=1|x)}{\sum_{x \in \mathcal{B}} \hat{p}(y=1|x)} \nonumber \\ 
&=& \frac{\sum_{x \in \mathcal{A}} \beta \cdot p(y=1|x, g=B)}{\sum_{x \in \mathcal{B}} \beta \cdot p(y=1|x, g=B)} \nonumber \\
  &<& \frac{\sum_{x \in \mathcal{A}} p(y=1|x, g=A)}{\sum_{x \in \mathcal{B}} p(y=1|x, g=B)} \nonumber \\ 
  &=& {\rho}_{A,B} \nonumber
\end{eqnarray}

so \texttt{PURPLE}'s estimate $\hat{\rho}_{A,B}$ provides a lower bound on the true relative prevalence ${\rho}_{A,B}$.

\paragraph{Empirical Validation} We verify this behavior empirically under synthetic violations of the  constant $p(y|x)$ assumption,  again using the synthetic data described in  \S \ref{sec:synth_data_generation}. Concretely, we vary the difference in the probability of a positive case between group $a$ and group $b$ ($p(y=1|x,g=a) - p(y=1|x,g=b)$), where $p(y=1|g=a) \geq p(y=1|g=b)$.

Fig. \ref{fig:supp_fig_cov_shift}B demonstrates how \texttt{PURPLE} consistently underestimates the true relative prevalence behavior empirically by plotting \texttt{PURPLE}'s behavior over a range of values for $\alpha$, where $\alpha \in [0,1]$ and $\alpha p(y=1|x,g=a) = p(y=1|x,g=b)$. In each case, \texttt{PURPLE} provides a lower bound on the true relative prevalence. We replicate this analysis for varying separations of the group-specific Gaussian distributions, where darker shades of purple correspond to group-specific distributions that are further from one another.

\subsection{Effect of Positive-Unlabeled Assumption Violations}
\label{sec:pu_violations}

\trackchange{Here we demonstrate how \texttt{PURPLE}'s performance varies over when the positive-unlabeled assumption is violated, meaning that the sample of observed positives contains some number of negatives. This can arise if the diagnostic used to create the set of observed positives exhibits a non-zero false positive rate. In Figure \ref{fig:supp_fig_violations}A, we plot the behavior of \texttt{PURPLE} as we increase the extent to  which the setting violates the PU assumption; specifically, we vary the percentage of the observed positives that truly are negative --- $p(y = 0|s = 1)$ --- from 0\% to 20\%. When $p(y=0|s=1) = 0$, this is equivalent to the positive-unlabeled assumption and \texttt{PURPLE} recovers the relative prevalence exactly, as expected. At greater violations of the PU assumption, \texttt{PURPLE}'s performance degrades somewhat, as expected.}

\subsection{Effect of Random-Diagnosis-Within-Groups Assumption Violations}
\label{sec:scar_violations}

\trackchange{We illustrate \texttt{PURPLE}'s behavior under violations of the random-diagnosis-within-groups assumption in Figure \ref{fig:supp_fig_violations}B. The assumption states that the probability that a true positive is diagnosed as such does not depend on $x$, or that $p(s=1|y=1,g,x) = p(s=1|y=1,g)$. We simulate violations of the assumption by generating $s$ according to $p(s=1|y=1,g,x) = c_g \cdot \sigma(\beta \cdot x_0)$. In other words, the diagnosis frequency for group $g$ is scaled by the sigmoid function of parameter $\beta$ multiplied by the first component of $x$. For $\beta = 0$, the setting adheres to the random-diagnosis-within-groups assumption. Higher values of $\beta$ translate to a higher correlation between the diagnosis probability and $x_0$. As Figure \ref{fig:supp_fig_violations}B demonstrates, \texttt{PURPLE} recovers the relative prevalence under no violation of the assumption ($\beta = 0$), and incurs small errors in the estimated relative prevalence as $\beta$ increases, as expected.}

\section{Datasets}
\label{sec:methods-datasets}

We make use of five datasets. We begin with two synthetic datasets: \emph{Gauss-Synth}, a completely synthetic dataset, and \emph{MIMIC-Semi-Synth}, a semi-synthetic dataset based on real health data. We then apply \texttt{PURPLE} to three non-synthetic datasets: \emph{MIMIC-IV  ED}, a dataset of electronic health records collected from a single hospital in the Boston area, and \emph{NEDS}, a dataset of emergency department visits occurring in the US in 2019.


We begin with synthetic and semi-synthetic data so that the ground truth labels $y$ are known, as is standard in the PU learning literature\cite{bekker2020learning}, enabling us to assess how well methods recover the relative prevalence. To assess performance of all methods in our synthetic and semi-synthetic experiments, we report the mean ratio of the estimated relative prevalence to the true relative prevalence over 5 random train/test splits of the dataset; values closer to 1 correspond to better performance. Code to reproduce all experiments can be found at \href{https://github.com/epierson9/invisible-conditions}{https://github.com/epierson9/invisible-conditions}.

\subsection{Gauss-Synth}
\label{sec:synth_data_generation}

We generate completely synthetic data by simulating group-specific features ($p(x|g)$) , and labels using a decision rule ($p(y|x)$). Formally, we simulate groups $a$ and $b$ using two 5D Gaussian distributions with different means: 

\begin{equation}
    x_i \thicksim
\begin{cases}
    \mathcal{N}_5(-\mathbf{1}, 16 \cdot \mathbf{1}) \text{ if } g_i = a\\
    \mathcal{N}_5(\mathbf{1}, 16 \cdot \mathbf{1}) \text{ if } g_i = b
\end{cases}
\end{equation}

The likelihood function ($p(y=1|x)$) is a logistic function of the signed distance to a hyperplane through the origin. Observed labels $s$ are drawn such that positive labels in group $i$ are observed with a probability of $c_{g_i}$. The generative model for $y$ and $s$ is:

\begin{eqnarray}
y_i &\thicksim&  \text{Bernoulli}(\sigma ( (\mathbf{1}^T x_i) \mathbin{/}  ||\mathbf{1}||)) 
\label{eqn:synth_y}\\
s_i  &\thicksim& \text{Bernoulli}(c_{g_i}y_i)
\end{eqnarray}

where $\sigma$ represents a logistic function ($\sigma(x) = \frac{1}{1 + e^{-x}}$) and $c_{g_i}$ is the group-specific diagnosis frequency for $g_i$. We draw 10,000 observations for group $a$ and 20,000 for group $b$. We create the separable data by modifying the generative model described above, which does not generate separable data. Specifically, we replace each  $p(y=1|x) > 0.5$ with  $p(y=1|x) = 1$ and each $p(y=1|x) < 0.5$ with $p(y=1|x) = 0$, and remove the 40\% of the data closest to the original decision boundary to ensure the classes are cleanly separable, as illustrated in Figure \ref{fig:separability_cov_shift}A.

\subsection{MIMIC-Semi-Synth} 
\label{sec:semi_synth}

We generate semi-synthetic data using MIMIC-IV, a public dataset of real patient visits to a Boston-area hospital  over the course of 2008-2018.\cite{johnson2020mimic} We filter out  ICD codes that appear 10 or fewer times, leaving 5,544 unique ICD codes. Each feature vector $x_i$ is a one-hot vector corresponding to the ICD codes assigned in a particular patient visit to the hospital. We generate true labels $y$ based on a set of suspicious symptoms. Formally, this replaces Equation (15) in our generative model with: 

\begin{equation}
    y_i \thicksim \text{Bernoulli}(\sigma(v_{sym}^T \mathbf{x_i}) \mathbin{/}  ||v_{sym}||) 
    \label{eqn:semisynth_y}
\end{equation}

where $v_{sym}$ is a one-hot encoding of the suspicious symptoms and $v_{sym}^T x_i$ corresponds to the number of suspicious symptoms present during a hospital visit. Thus, the probability a patient has a medical condition is a logistic function of the number of suspicious symptoms. As before, we have $s_i \thicksim \text{Bernoulli}(c_{g_i}y_i)$. 

In all experiments, we compute the relative prevalence for Black (group $a$) versus white (group $b$) patients since these are the largest race groups in MIMIC data. We filter the dataset for patients belonging to each group. However, our method can be applied to more than 2 groups, as described above. To assess how our method performs under diverse conditions, we experiment with selecting the suspicious symptoms $v_{sym}$ in four different ways:

\paragraph{Common Symptoms} \label{sec:common_simulation_setup}We identify the 50 most common ICD codes in MIMIC-IV and randomly select 25 to be suspicious symptoms (Table \ref{tab:common_symptoms}). Group $a$ consists of 73,090 visits from Black patients ($p(y=1|g=a) = 0.157$) and group $b$ consists of $305,002$ visits from White patients ($p(y=1|g=b) = 0.185$).

\paragraph{High Relative Prevalence Symptoms} \label{sec:high_rp_simulation_setup} We filter out ICD codes that appear fewer than 50 times in each group and patients less than 18 years old. After ranking the ICD codes by relative prevalence---prevalence among visits by white patients, divided by prevalence among visits by Black patients---we select the top 10 ICD codes as our suspicious symptom set (Table \ref{tab:high_rp_symptoms}). Group $a$ contains 14,618  visits from Black patients $p(y=1|g=a) = 0.061$) and group $b$ contains 61,000 visits from white patients ($p(y=1|g=b) = 0.098$).

\paragraph{Correlated Symptoms} \label{sec:corr_simulation_setup}We consider endometriosis, a widely under-diagnosed condition\cite{moradi2014impact}.  We define our suspicious symptoms as the symptoms most highly associated with known endometriosis codes. We first identify a set of patients who receive any one of 10 gold-standard ICD endometriosis diagnosis codes (Table \ref{tab:endo_diagnoses}). We then identify the ICD codes which are most highly associated with a gold-standard diagnosis of endometriosis: for each ICD code, we compute the ratio $\frac{\text{prevalence of ICD code among endometriosis patients}}{\text{prevalence of ICD code among all patients}}$. We define our suspicious symptoms as the 25 ICD codes with the highest value of this ratio, which includes known endometriosis symptoms such as ``Excessive and frequent menstruation with regular cycle" and ``Pelvic and perineal pain" (Table \ref{tab:endo_symptoms}). We do not include the 10 ICD codes used to determine the 25 suspicious symptoms in $x$. We filter for female patients because endometriosis is extremely rare among male patients\cite{jabr2014unusual}, leaving 47,138 unique hospital visits from Black patients ($p(y=1|g=a) = 0.0534$) and 165,653 unique hospital visits from white patients ($p(y=1|g=a) = 0.0495$).

\paragraph{Recognized Symptoms for IPV} \label{sec:recognized_simulation_setup}Prior work has found that suspicious symptoms for IPV include head, neck and facial injuries\cite{wu2010pattern}. The symptoms in this experiment consist of the 100 ICD codes corresponding to these injuries (Table \ref{tab:ipv_symptoms}). We filter for female patients because the symptoms associated with IPV in male patients are not well understood\cite{houry2008differences}. We also filter out patients less than 18 years old because it is difficult to distinguish between intimate partner violence and child abuse in minors. This results in a dataset with $p(y=1|g=a) = 0.0541$ (25,546 unique patient visits) and $p(y=1|g=b) = 0.0568$ (80,227 unique patient visits). 

In Sections \ref{sec:mimic-iv-ed} and \ref{sec:NEDD} we describe construction of the two non-synthetic datasets, MIMIC-IV and NEDS. In all these datasets, $y$ is unknown, so we need only define the features $x$ and known positive examples in which $s=1$. 


\subsection{MIMIC-IV ED}
\label{sec:mimic-iv-ed}

\paragraph{Data Filtering} MIMIC-IV contains two related databases: one representing diagnoses made in the hospital (which we will refer to as the hospital database) and one representing diagnoses made in the emergency department. There are slight inconsistencies between the two, as is to be expected; for example, one emergency department stay can be associated with multiple unique hospital admissions. We exclude 600 hospital admissions linked to multiple emergency department stays, and exclude 59 emergency department stays associated with invalid hospital admissions. For emergency department stays that result in a hospital admission, we include all diagnoses assigned in the emergency department or hospital. We also include patient visits that appear only in the hospital database, but indicate admission through the emergency department via the ``admission\_location" field. We provide code to replicate these preprocessing steps.

We further filter for patients who are female and above 18. We do so because we are interested in the relative prevalence of intimate partner violence between subgroups of adult female patients.  This leaves 293,297 individual hospital visits over 133,470 unique patients. For each demographic attribute we wish to analyze disparities over (ethnicity, insurance status, and marital status), we also filter out patients who are missing data for this attribute. This translates to 192,768 stays across marital statuses, 208,512 stays across ethnicities, and 108,948 stays across insurance statuses. We have significantly fewer stays in the insurance subgroups because we only patients for whom the insurance status is known (i.e. Medicare and Medicaid recipients). We produce 5 randomized dataset splits, where we reserve 60\% of the data for training, 20\% for validation, and 20\% for testing. Each patient appears in only one of these sets. 

\paragraph{Defining features $x$} We represent each patient visit as a one-hot encoding of the ICD codes assigned. Concretely, 15,699 features represent each patient visit, where each feature corresponds to the presence or absence of one ICD code (across the ICD-9 and ICD-10 standards). Note that this is different from the semi-synthetic set-up and we do \emph{not} filter out codes that appeared fewer than 10 times. We do this because IPV itself is rare, and we do not want to exclude symptoms which are predictive of IPV and do not occur frequently.

\paragraph{Defining $s=1$} To define examples where $s=1$ (known positive examples) we use criteria for reported instances of intimate partner violence from prior work. The most specific code is E967.3, or "Battering by an intimate partner", drawn from the ICD-9 standard. Other codes include V6111, or "Counseling for victim of spousal or partner abuse". The full code set can be found in Table \ref{tab:ipv_diagnoses}. If a patient receives any one of the the codes in the positive code set, the visit is deemed to be positively labeled for intimate partner violence.

\paragraph{Defining $g$} We define the groups over which we quantify disparities via demographic variables associated with each electronic health record. These include race/ethnicity (Black, white, Asian, Hispanic/Latino), insurance status (Medicare or Medicaid), and marital status (Legally Unmarried, Married, Divorced). 

\subsection{NEDS}
\label{sec:NEDD}
The National Emergency Department Sample (NEDS) is the largest publicly available, all-payer database describing visits to hospital-owned emergency departments in the US, and is commonly used in studies of disease prevalence. In this work, we make use of NEDS 2019. The survey from this year produced a dataset containing 33.1 million visits, which represent the 143 million total visits occuring in hospital-owned EDs across the US in 2019. To create nationally representative estimates, NEDS releases ``discharge weights", which allow the analyst to reweight estimates to represent the universe of emergency department visits. We follow this procedure to reweight our relative prevalence estimates.

\paragraph{Data Filtering} As with MIMIC-IV, we include visits from patients who are female and above the age of 18. The resulting dataset contains 15,357,528 visits.

\paragraph{Defining features $x$}  As before, we treat the ICD codes logged during a visit as the input features, as a proxy for the symptoms a patient presents with. The feature set consists of 19,710 ICD-10 codes (excluding those used to identify positive cases, as described in the next section), which we one-hot encode to create features for each ED visit.

\paragraph{Defining $s = 1$} We use the same criteria to identify positive cases as with MIMIC-IV: we consider a case to be labeled positive if it is associated with  if it has any of the  codes in Table \ref{tab:ipv_diagnoses}.

\paragraph{Defining $g$} We define groups according to race/ethnicity groups, insurance status, income quartile, and urban/rural designation. Income quartile is calculated using the estimated median household income in the patient's zipcode. In 2019, the first quartile corresponds to an estimated median houshold income between \$0 and \$48k, the second to \$48k to \$61k, the third to \$61k - \$82k, and the fourth to the remaining zipcodes. The urban/rural designations are based on the population of the patient's home county. From most to least rural, the four categories are: 50k residents  to 250k residents, 250k residents to 1 million residents, 1 million residents in a fringe metropolitan county, and 1 million residents in a central metropolitan county.

\section{Baselines}
\label{sec:baselines}

Each of the baseline methods described below is designed to estimate the \emph{absolute} prevalence. To obtain the relative prevalence, we apply each baseline to groups $a$ and $b$ individually to obtain estimates of the absolute prevalence in each group; we then divide the resulting quantities to produce an estimate of the relative prevalence. To provide consistent comparisons, we constrain each baseline to the use same function class as \texttt{PURPLE}. We do not compare to baselines in the epidemiology literature because they assume access to external information (e.g. diagnostic accuracy) that is often not available; we also do not consider work that places parametric assumptions on $p(y=1|x)$\cite{lazkecka2021estimating,jaskie2019positive,teisseyre2020different,furmanczyk2022joint} because these assumptions will not hold in general.


\begin{itemize}
    \itemsep 0em
    \item \emph{Negative}: Assigns all unlabeled examples a negative label. This approach replaces $p(y=1|x)$ with $p(s=1|x)$ and assumes no underreporting occurs. Past work refers to this model as a \emph{nontraditional classifier}\cite{elkan2008learning} (NTC). We use \emph{sklearn}'s logistic regression implementation with no regularization and default settings for all other hyperparameters, trained with target $s$.
    \item \emph{KM2}: Models the distribution of unlabeled examples as a mixture of the positive and negative distribution and estimates the proportion of positives using a kernel mean embedding approach\cite{ramaswamy2016mixture}. This method is known to perform poorly on large datasets with many features\cite{ivanov2020dedpul}. KM2 assumes that there exists a function that only selects positive examples.
    \item \emph{DEDPUL}: Uses an NTC to map each example to a predicted probability of being labeled, and performs mixture proportion estimation using the classifier's outputs on the unlabeled sample\cite{ivanov2020dedpul}. Specifically, the method applies heuristics to the estimated densities of the positive and unlabeled  distribution.
    \item \emph{BBE}: Identifies a small subset of positive examples using the outputs of an NTC on the positive and unlabeled sample\cite{garg2021mixture}. The method uses this subset to infer the proportion of positive examples in the unlabeled sample.
    \item \emph{Oracle}: Uses the true label $y$ to estimate $p(y=1|x)$. Importantly, this method cannot actually be applied in real data since $y$ is unobserved, but it represents an upper bound on performance.
\end{itemize}

\clearpage


\linespread{1.0}\selectfont
\FloatBarrier
\renewcommand\thesection{S\arabic{section}}
\renewcommand\thefigure{S\arabic{figure}}
\renewcommand\thetable{S\arabic{table}}
\renewcommand{\figurename}{Figure}
\renewcommand{\tablename}{Table}
\setcounter{section}{0}
\setcounter{figure}{0}
\setcounter{table}{0}
\section{Related Work}
\label{sec:related_work}
Because medical underdiagnosis is so ubiquitous, past work in both epidemiology and machine learning has attempted to model and correct for it. 

\paragraph{Epidemiology} Numerous works in epidemiology address prevalence estimation in the context of untrustworthy diagnostic tests\cite{lash2014good} (a type of ascertainment bias, where diagnostic tests do not reliably distinguish positive and negative cases\cite{delgado2004bias}). Early work characterized the impact of test sensitivity, test specificity, and underlying disease prevalence on the accuracy of relative prevalence estimates\cite{copeland1977bias,greenland1983correcting,walter1988estimation,kristensen1992bias,neuhaus1999bias,reiczigel2017exact}, ultimately developing methods to adjust relative risk estimates given test characteristics\cite{magder1997logistic,diggle2011estimating,lewis2012tutorial,haine2018selection}. The field naturally moved to consider the utility of tests with unknown specificity and sensitivity, and proposed methods to infer these test characteristics. Common approaches include using external information (for example, ground truth annotations\cite{lyles2011validation,sekar1949method,simeone1995plan,hay2003estimating,mckeganey1992female}, multiple tests\cite{hui1980estimating,walter1988estimation, pepe2007insights},  or informative priors\cite{lewis2012association}). However, this type of information is frequently unavailable for underreported medical conditions. We often have no access to ground truth, only a single diagnosis per patient, and no notion of how accurate that diagnosis truly is. 

\paragraph{Machine Learning} The machine learning literature has also modeled underdiagnosis using the positive-unlabeled (PU) learning framework, which assumes that only some positive cases are correctly labeled as positive, and the unlabeled examples consist of both true negatives and unlabeled true positives. 
A key result in PU learning proves that without further assumptions --- e.g., about the structure of the positive and negative distributions --- it is impossible to estimate the prevalence of an underreported condition\cite{scott2015rate}. Many PU learning methods consequently introduce assumptions about the structure of the positive and negative distributions in order to recover the prevalence, including complete separability, positive subdomain\cite{elkan2008learning,scott2004multi,christoffel2016class,liu2015classification,bekker2018estimating}, positive function\cite{ramaswamy2016mixture,ivanov2020dedpul,garg2021mixture}, and irreducibility\cite{blanchard2010semi,jain2016estimating}. All of these conditions represent different ways of encoding the basic assumption that the negative distribution does not fully contain the positive distribution: in other words, that there is a region of the feature space where cases are certain to be true positives. However, this is a restrictive assumption which, while potentially suitable in other PU settings, is unlikely to hold in health data\cite{chen2017machine} because it is rare that a set of symptoms corresponds to a positive diagnosis with 100\% certainty. This is especially true in women's health, where symptoms are frequently not specific to a particular condition (for example, menstrual irregularities could suggest a number of underlying conditions, rather than just one).

\begin{table}[htb]
\centering
\resizebox{.5\columnwidth}{!}{%
\begin{tabular}{ll}
\toprule
ICD Code &                               ICD Code Description \\
\midrule
    2724 &               Other and unspecified hyperlipidemia \\
    2859 &                                Anemia, unspecified \\
   30000 &                         Anxiety state, unspecified \\
     412 &                          Old myocardial infarction \\
   41401 & Coronary atherosclerosis of native coronary artery \\
   42731 &                                Atrial fibrillation \\
     496 & Chronic airway obstruction, not elsewhere class... \\
    5849 &                  Acute kidney failure, unspecified \\
    E039 &                        Hypothyroidism, unspecified \\
    E669 &                               Obesity, unspecified \\
    F419 &                      Anxiety disorder, unspecified \\
     I10 &                   Essential (primary) hypertension \\
   I4891 &                    Unspecified atrial fibrillation \\
    N179 &                  Acute kidney failure, unspecified \\
   V5861 &          Long-term (current) use of anticoagulants \\
   V5866 &                 Long-term (current) use of aspirin \\
    Y929 &                Unspecified place or not applicable \\
   Z7901 &          Long term (current) use of anticoagulants \\
    Z794 &                 Long term (current) use of insulin \\
\bottomrule
\end{tabular}
}
\caption{\textbf{Common Suspicious Symptoms.} Suspicious symptoms for the semi-synthetic dataset created using common symptoms. We select these symptoms randomly from the 50 most common ICD codes in MIMIC-IV.}
\label{tab:common_symptoms}
\end{table}

\begin{table*}[htb]
\centering
\resizebox{.5\columnwidth}{!}{%
\begin{tabular}{ll}
\toprule
ICD Code &                               ICD Code Description \\
\midrule
   20410 &  Chronic lymphoid leukemia, without mention of ... \\
   41402 &  Coronary atherosclerosis of autologous vein by... \\
   43310 &  Occlusion and stenosis of carotid artery witho... \\
   53085 &                                Barrett's esophagus \\
    5569 &                    Ulcerative colitis, unspecified \\
    8730 &  Open wound of scalp, without mention of compli... \\
   K2270 &              Barrett's esophagus without dysplasia \\
   V1051 &  Personal history of malignant neoplasm of bladder \\
    V422 &                 Heart valve replaced by transplant \\
   V4581 &                        Aortocoronary bypass status \\
\bottomrule
\end{tabular}}
\caption{\textbf{High Relative Prevalence Suspicious Symptoms.} Diagnoses that are disproportionately more common among white patients. We filter out codes that appear fewer than 50 times in either group (white or Black patients). We then rank all codes by their prevalence among white patients divided by prevalence among Black patients. We then select the top 10 ICD codes. Our motivation is to ensure \texttt{PURPLE} recovers accurate relative prevalence estimates when the suspicious symptoms are highly correlated with the group variable, and when the true relative prevalence estimate is far from 1.}
\label{tab:high_rp_symptoms}
\end{table*}
\begin{table}[htb]
\centering
\resizebox{.5\columnwidth}{!}{%
\begin{tabular}{ll}
\toprule
ICD Code &                            ICD Code Description \\
\midrule
     N80 &                                   Endometriosis \\
    N800 &                         Endometriosis of uterus \\
    N801 &                          Endometriosis of ovary \\
    N802 &                 Endometriosis of fallopian tube \\
    N803 &              Endometriosis of pelvic peritoneum \\
    N804 & Endometriosis of rectovaginal septum and vagina \\
    N805 &                      Endometriosis of intestine \\
    N806 &                 Endometriosis in cutaneous scar \\
    N808 &                             Other endometriosis \\
    N809 &                      Endometriosis, unspecified \\
    6179 &                 Endometriosis, site unspecified \\
\bottomrule
\end{tabular}
}
\caption{\textbf{Endometriosis Diagnoses.} Diagnoses used  to identify endometriosis cases. We use all symptoms containing reference to endometriosis, and identify these by filtering for ICD codes whose long descriptions (as described by MIMIC-IV\cite{johnson2020mimic}) contain the word ``endometriosis".}
\label{tab:endo_diagnoses}
\end{table}

\begin{table}[htb]
\centering
\resizebox{.5\columnwidth}{!}{%
\begin{tabular}{ll}
\toprule
ICD Code &                               ICD Code Description \\
\midrule
   33819 &                                   Other acute pain \\
    5951 &                      Chronic interstitial cystitis \\
    6205 & Torsion of ovary, ovarian pedicle, or fallopian... \\
    6260 &                            Absence of menstruation \\
   78902 &                Abdominal pain, left upper quadrant \\
   78904 &                Abdominal pain, left lower quadrant \\
   78905 &                        Abdominal pain, periumbilic \\
    7891 &                                       Hepatomegaly \\
    C561 &                  Malignant neoplasm of right ovary \\
    D251 &                     Intramural leiomyoma of uterus \\
    D252 &                     Subserosal leiomyoma of uterus \\
    D259 &                   Leiomyoma of uterus, unspecified \\
    D270 &                     Benign neoplasm of right ovary \\
     E43 &    Unspecified severe protein-calorie malnutrition \\
    F911 &             Conduct disorder, childhood-onset type \\
   G8921 &                         Chronic pain due to trauma \\
    K661 &                                     Hemoperitoneum \\
    N739 &    Female pelvic inflammatory disease, unspecified \\
    N952 &                  Postmenopausal atrophic vaginitis \\
    O210 &                        Mild hyperemesis gravidarum \\
   O2341 & Unspecified infection of urinary tract in pregn... \\
  O26891 & Other specified pregnancy related conditions, f... \\
    Q600 &                         Renal agenesis, unilateral \\
   R1310 &                             Dysphagia, unspecified \\
     R17 &                               Unspecified jaundice \\
\bottomrule
\end{tabular}
}
\caption{\textbf{Correlated  Suspicious Symptoms.} Suspicious symptoms for the semi-synthetic dataset created using symptoms correlated with endometriosis. We select the 25 ICD codes with the highest relative proportion among endometriosis patients (where endometriosis patients are identified as patients receiving any ICD code appearing in Table \ref{tab:endo_diagnoses}).}
\label{tab:endo_symptoms}
\end{table}
\begin{table*}[htb]
\centering
\resizebox{.5\columnwidth}{!}{%
\begin{tabular}{ll}
\toprule
ICD Code &                               ICD Code Description \\
\midrule
    7842 &           Swelling, mass, or lump in head and neck \\
    9100 & Abrasion or friction burn of face, neck, and sc... \\
     920 &   Contusion of face, scalp, and neck except eye(s) \\
   95901 &                           Head injury, unspecified \\
 S0003XA &              Contusion of scalp, initial encounter \\
 S0011XA & Contusion of right eyelid and periocular area, ... \\
 S0012XA & Contusion of left eyelid and periocular area, i... \\
 S0990XA &      Unspecified injury of head, initial encounter \\
\bottomrule
\end{tabular}}
\caption{\textbf{Recognized  Suspicious Symptoms.} Diagnoses associated with intimate partner violence. We first identify all ICD codes that refer to head, neck, and face injuries (100 ICD  codes). We filter out codes that appear fewer than 10 times as a part of dataset preprocessing (\S\ref{sec:semi_synth}), which leaves the 8 codes listed here.}
\label{tab:ipv_symptoms}
\end{table*}
\begin{table}[htb]
\centering
\resizebox{.5\columnwidth}{!}{%
\begin{tabular}{ll}
\toprule
ICD Code &                               ICD Code Description \\
\midrule
   E9672 &  Perpetrator of child and adult abuse, by mothe... \\
   E9673 &  Perpetrator of child and adult abuse, by spous... \\
   E9671 &  Perpetrator of child and adult abuse, by other... \\
   E9670 &  Perpetrator of child and adult abuse, by fathe... \\
   E9679 &  Perpetrator of child and adult abuse, by unspe... \\
   V6111 &  Counseling for victim of spousal and partner a... \\
   99581 &                               Adult physical abuse \\
   99585 &                      Other adult abuse and neglect \\
 T7411XA &  Adult physical abuse, confirmed, initial encou... \\
 T7411XD &  Adult physical abuse, confirmed, subsequent en... \\
 T7411XS &           Adult physical abuse, confirmed, sequela \\
   99580 &                    Adult maltreatment, unspecified \\
   T7611 &                    Adult physical abuse, suspected \\
 T7611XA &  Adult physical abuse, suspected, initial encou... \\
 T7611XD &  Adult physical abuse, suspected, subsequent en... \\
 T7611XS &           Adult physical abuse, suspected, sequela \\
 T7611XA &  Adult physical abuse, suspected, initial encou... \\
 T7611XD &  Adult physical abuse, suspected, subsequent en... \\
 T7611XS &           Adult physical abuse, suspected, sequela \\
    Y070 &  Spouse or partner, perpetrator of maltreatment... \\
   Y0701 &   Husband, perpetrator of maltreatment and neglect \\
   Y0702 &      Wife, perpetrator of maltreatment and neglect \\
   Y0703 &  Male partner, perpetrator of maltreatment and ... \\
   Y0704 &  Female partner, perpetrator of maltreatment an... \\
    Y079 &  Unspecified perpetrator of maltreatment and ne... \\
\bottomrule
\end{tabular}
}
\caption{\textbf{IPV Diagnoses.} Diagnoses used  to identify positive intimate partner violence cases. We use symptoms collected from prior work, including surveys used to study the prevalence of intimate partner violence\cite{leone2019social,davidov2015united,schafer2008using,halpern2006protocol}, in addition to manual examination of the ICD-10 and ICD-9 codebook.}
\label{tab:ipv_diagnoses}
\end{table}

\FloatBarrier
\begin{figure}[t!]
  \centering
  \includegraphics[width=\columnwidth]{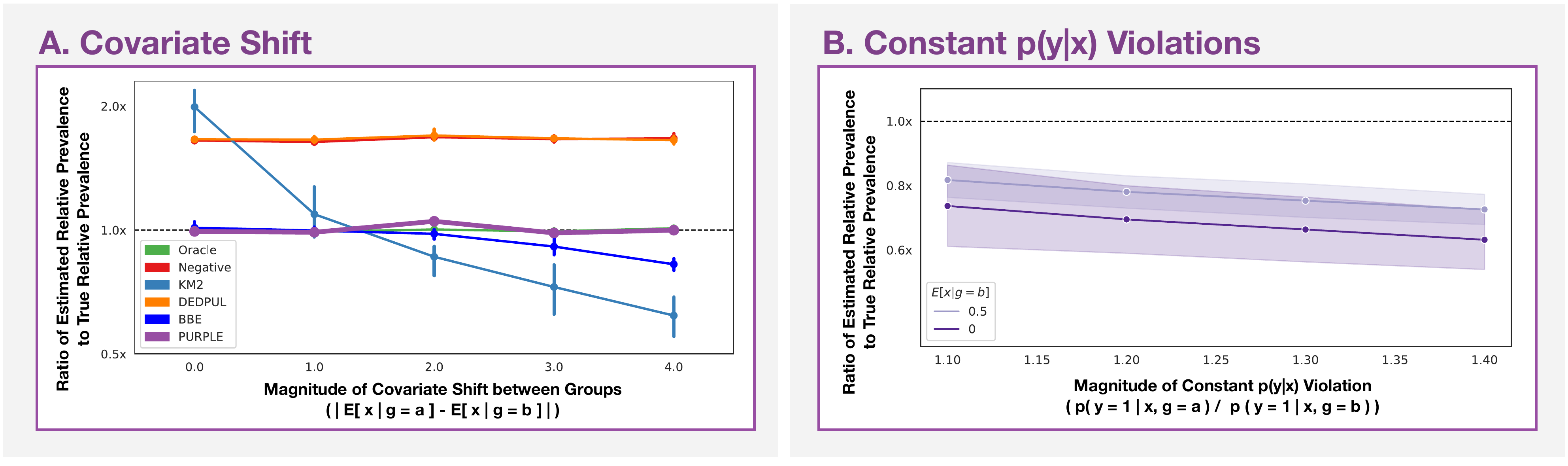}
  \caption{\text{Robustness checks on synthetic data.} \textbf{A) \texttt{PURPLE} is robust to differences in group-specific covariate distributions.} \texttt{PURPLE} produces consistently accurate relative prevalence estimates across simulations with increasingly different groups, where we simulate group differences by altering the magnitude of the difference between group means ($|\text{E}[x|g=a] - \text{E}[x|g=b]|$). \textbf{B) \texttt{PURPLE} still yields useful lower bounds on prevalence disparities even if $p(y=1|x)$ varies beween groups.}  Under a plausible violation of the assumption that $p(y=1|x)$ remains constant across groups, \texttt{PURPLE} can lower-bound the true relative prevalence. Specifically, if the same symptoms correspond to a higher likelihood of a condition in the group with a higher overall prevalence (i.e., $p(y = 1 | x, g = a) > p(y = 1 | x, g = b)$ if $p(y=1|g=a) > p(y=1|g=b)$), \texttt{PURPLE} produces an underestimate of the true relative prevalence. Different shades of purple correspond to different magnitudes of covariate shift between groups.}
  \label{fig:supp_fig_cov_shift}
\end{figure}

\begin{figure}[t!]
  \centering
  \includegraphics[width=\columnwidth]{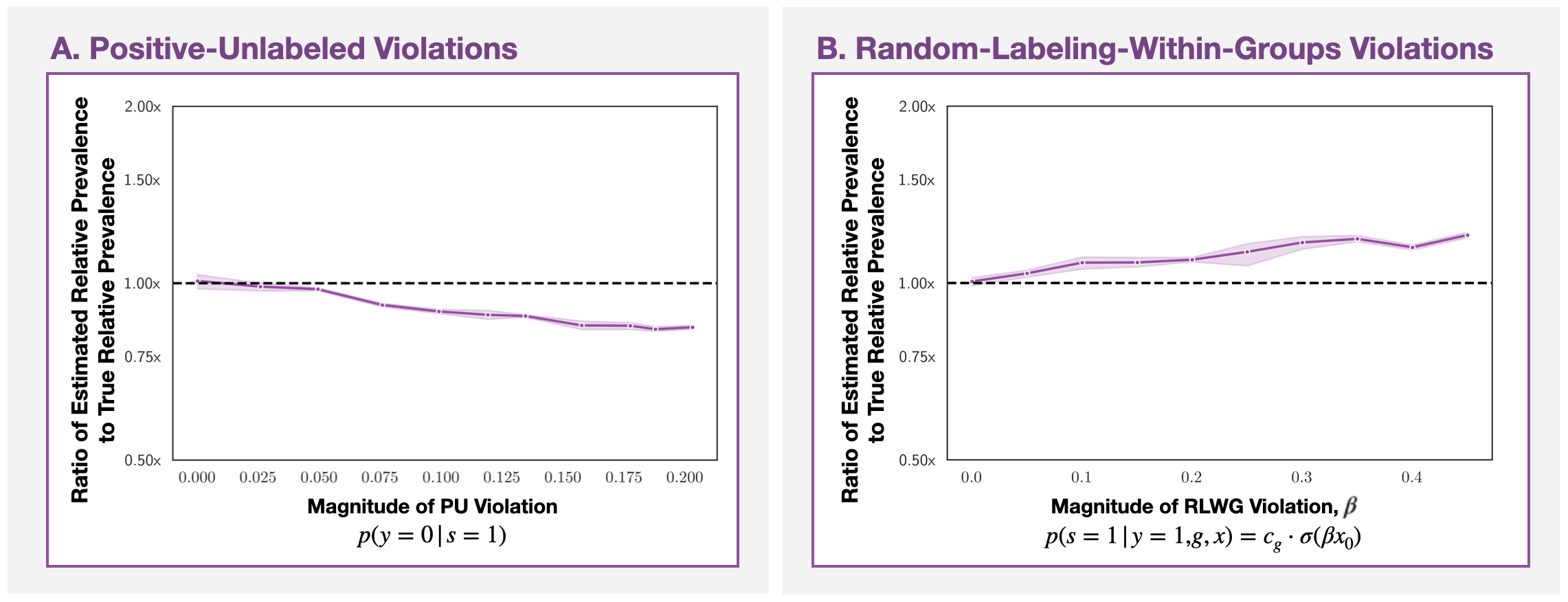}
  \caption{\trackchange{\text{Robustness to violations of assumptions on synthetic data.} \textbf{A) Effects of violations of positive-unlabeled assumption.} We plot the behavior of \texttt{PURPLE} as we increase the percentage of false positives present in the ``positive sample'', or all examples where $s=1$. As expected, \texttt{PURPLE} recovers the relative prevalence exactly when there are no false positives in the positive sample, and performance degrades as this percentage increases. \textbf{B) Effects of violations of the random-diagnosis-within-groups assumption.} As the magnitude of $\beta$ increases, the diagnosis probability becomes more correlated with feature $X_0$, thereby violating the random-diagnosis-within-groups assumption. As expected, these violations, produce larger errors in \texttt{PURPLE}'s estimates of the relative prevalence. We illustrate the impact of these violations to characterize \texttt{PURPLE}'s behavior under violations of the central assumptions and provide checks for whether these assumptions are collectively violated in Sec. \ref{sec:assumption_checks}.}}
  \label{fig:supp_fig_violations}
\end{figure}


\begin{figure}[t!]
  \centering
  \includegraphics[width=\columnwidth]{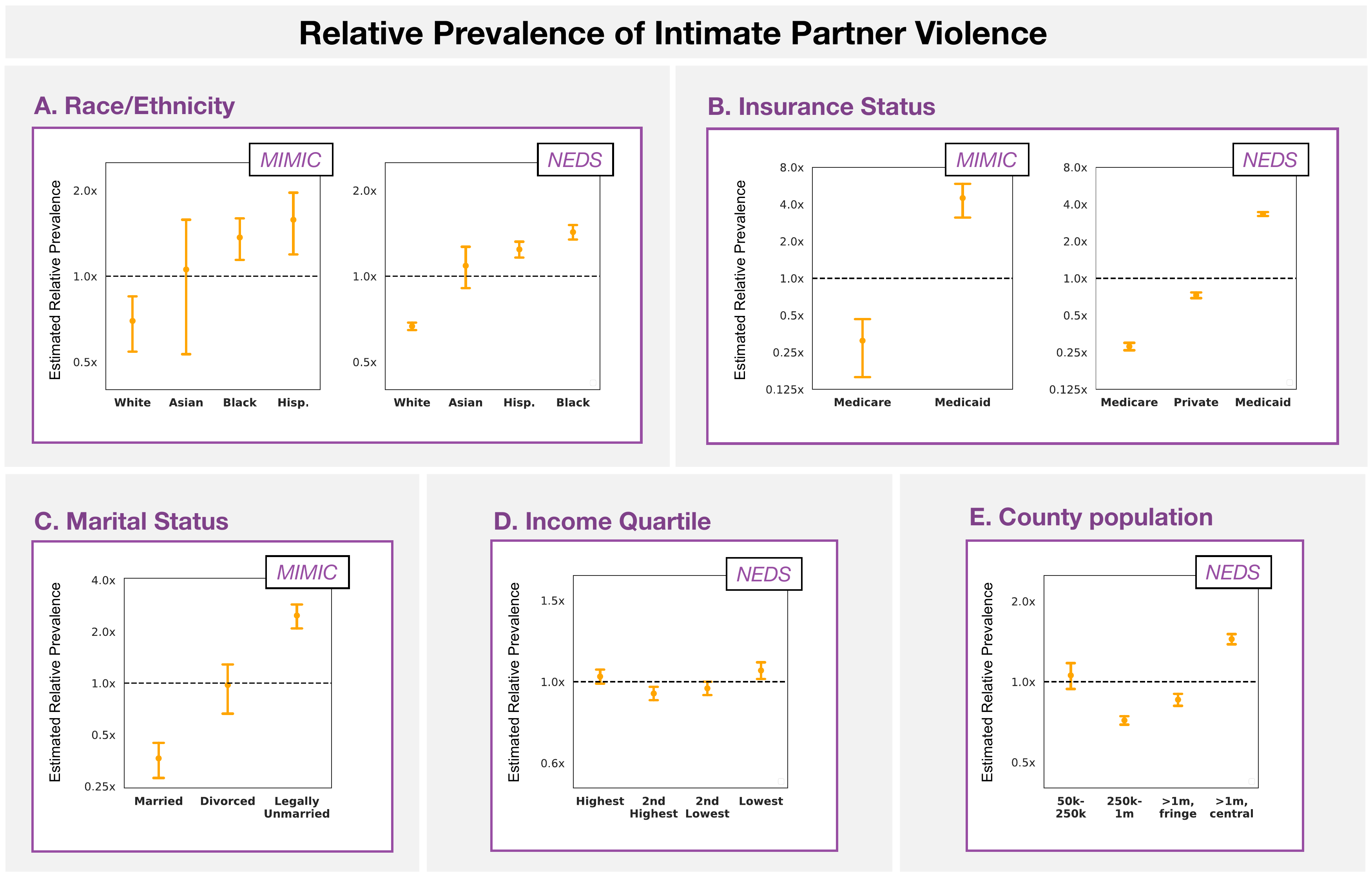}
  \caption{\textbf{Comparing \texttt{PURPLE} to relative prevalence estimates which do not correct for underdiagnosis.} We compare \texttt{PURPLE}'s relative prevalence estimates to relative prevalence estimates based on $p(s=1|g)$ across groups --- which simply uses observed diagnoses $s$, and does not correct for underdiagnosis. Orderings across groups for race, insurance status, and marital status are qualitatively similar for both methods. When we look at intimate partner violence prevalence across income quartiles, however, we see that \texttt{PURPLE}'s relative prevalence estimates agree more closely with prior work, which has found that rates of intimate partner violence decrease with income. Further, \texttt{PURPLE} infers that intimate partner violence is underdiagnosed among women in the lowest income quartile, which is supported by prior work on underdiagnosis among low-income patients. In contrast, the estimates which do not correct for underdiagnosis do not reveal a monotonic trend, and are harder to reconcile with past work.}
  \label{fig:observed_rps}
\end{figure}


\begin{figure}[t!]
  \centering
  \includegraphics[width=\columnwidth]{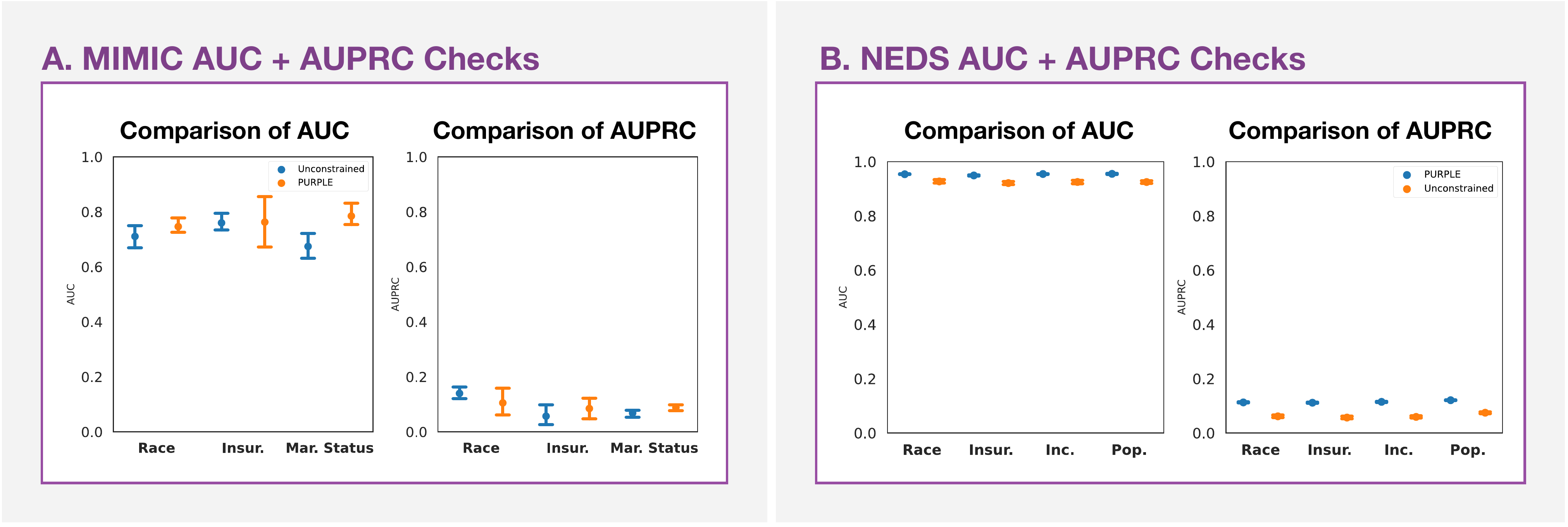}
  \caption{\textbf{Model check: Comparison of \texttt{PURPLE} and unconstrained model on MIMIC-IV data (A) and NEDS data (B).} The unconstrained model is a model with interactions between group status and features, allowing it to learn different likelihood functions $p(y=1|x)$ for each group, in contrast to \texttt{PURPLE}, which assumes  that the likelihood function remains constant across groups. We report results of the unconstrained model for each demographic category individually because each category corresponds to different subsets of patients (for example, the insurance category contains patients who are on Medicare or Medicaid, which filters out patients who are not on either). AUCs and AUPRCs of \texttt{PURPLE} are similar to the unconstrained model for all demographic categories and both datasets, indicating that \texttt{PURPLE}'s restriction in model form does not harm performance. (In some cases, \texttt{PURPLE} actually does slightly better than the unconstrained model, likely indicating that the interaction terms the unconstrained model has access to provide noise but not signal, making it more vulnerable to overfitting).}
  \label{fig:auc_auprc_check}
\end{figure} 


\begin{figure}[t!]
  \centering
  \includegraphics[width=\columnwidth]{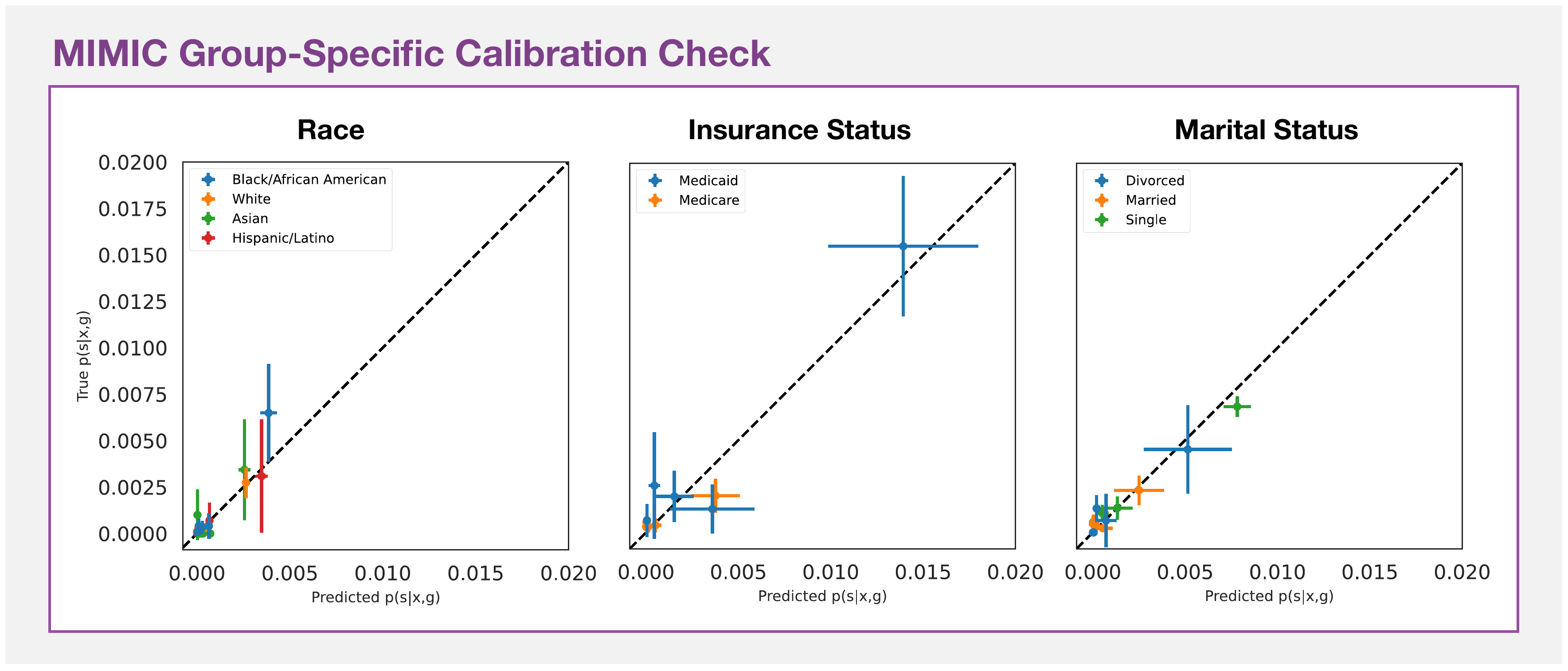}
  \caption{\textbf{Model check: Group-specific calibration in MIMIC-IV data.}  \texttt{PURPLE} outputs calibrated probabilities across subgroups. We bin predictions into quintiles  and plot the means of $\hat{p}(s=1|x,g)$ and $p(s=1|x,g)$ over 5 randomized dataset splits; errorbars show standard deviations across the 5 splits. The width of the error bars is due to the small number of positive examples in the intimate partner violence data. To ensure our calibration estimates are not overly noisy, in this analysis only we omit groups with fewer than 500 health records in the test set (e.g., Hispanic patients).}
  \label{fig:calibration_check_mimic}
\end{figure}


\begin{figure}[t!]
  \centering
  \includegraphics[width=\columnwidth]{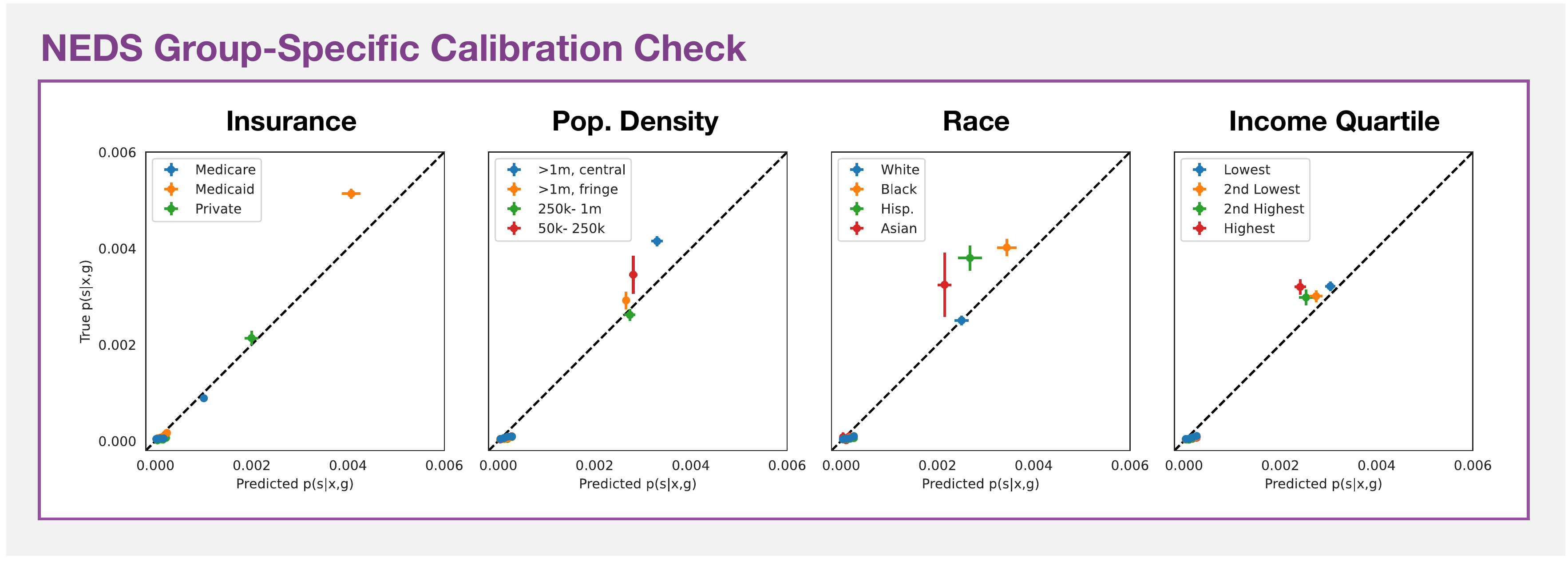}
  \caption{\textbf{Group-specific calibration in NEDS.} \texttt{PURPLE} is well-calibrated across subgroups. We bin predictions on the test set into quintiles and compare the average predicted probability of diagnosis ($\hat{p}(s=1|x,g)$, horizontal axis) and the true probability of diagnosis ($p(s=1|x,g)$, vertical axis) within each bin. We produce error bars, as before, by plotting the standard deviation of both $\hat{p}(s=1|x,g)$ and $p(s=1|x,g)$ for each bin across 5 randomized test sets.}
  \label{fig:calibration_check_neds}
\end{figure}

\FloatBarrier

\FloatBarrier
\clearpage

\end{document}